\documentclass{aa}  
\pdfoutput=1
\usepackage{graphicx}
\usepackage{float}
\usepackage{multirow}
\usepackage{subcaption} 
\usepackage{lscape}
\usepackage{txfonts}
\usepackage{xcolor}
\usepackage{placeins}
\usepackage[hidelinks]{hyperref}
\newcommand{\DEM}{\text{DEM}}

\newcommand{\beq}{\begin{equation}}
\newcommand{\eeq}{\end{equation}}

\newcommand{\bal}{\begin{align}}
\newcommand{\eal}{\end{align}}
\newcommand{\mltc}[1]{\multicolumn{1}{c}{#1}}

\newcommand\LF{\ensuremath{\textrm{LF}}}
\newcommand\HF{\ensuremath{\textrm{HF}}}
\newcommand\phot{\ensuremath{\textrm{P}}}
\newcommand\coro{\ensuremath{\textrm{C}}}

\begin{document}

   \title{SPICE Connection Mosaics to link the Sun's surface and the heliosphere}

\author{T. Varesano\inst{1,2,3}
\and D. M. Hassler\inst{2}
\and N. Zambrana Prado\inst{4}
\and J. Plowman\inst{2}
\and G. Del Zanna\inst{5}
\and S. Parenti\inst{6}
\and H. E. Mason \inst{5}
\and A. Giunta\inst{7}
\and F. Auchère\inst{6}
\and M. Carlsson\inst{8}
\and A. Fludra\inst{7}
\and H. Peter\inst{9}
\and D. Müller\inst{10}
\and D. Williams\inst{11}
\and R. Aznar Cuadrado\inst{9}
\and K. Barczynski\inst{12}
\and E. Buchlin\inst{6}
\and M. Caldwell\inst{7}
\and T. Fredvik\inst{8}
\and T. Grundy\inst{7}
\and S. Guest\inst{7}
\and L. Harra\inst{13}
\and M. Janvier\inst{6}
\and T. Kucera\inst{4}
\and S. Leeks\inst{7}
\and W. Schmutz\inst{12}
\and U. Schuehle\inst{9}
\and S. Sidher\inst{7}
\and L. Teriaca\inst{9}
\and W. Thompson\inst{14}
\and S. L. Yardley\inst{15,16,17}
}

\institute{
Institut National Polytechnique de Grenoble, 38000 Grenoble, France
\and
Southwest Research Institute, Boulder, CO 80302, USA
\and Department of Aerospace Engineering Sciences, University of Colorado Boulder, Boulder, CO, USA
\and
NASA Goddard Space Flight Center, Greenbelt, MD, USA
\and DAMTP, Centre for Mathematical Sciences, Wilberforce Road, Cambridge CB3 0WA, UK
\and
Universite Paris-Saclay, CNRS, Institut d’Astrophysique Spatiale, 91405, Orsay, France
\and
RAL Space, UKRI STFC Rutherford Appleton Laboratory, Didcot, United Kingdom
\and
Institute of Theoretical Astrophysics, University of Oslo, Norway
\and
Max-Planck-Institut fur Sonnensystemforschung, Gottingen, Germany
\and
European Space Agency, ESTEC, Noordwijk, The Netherlands
\and
European Space Agency, ESAC, Villanueva de la Canada, Spain
\and
Physikalisch-Meteorologisches Observatorium Davos, World Radiation Center, Davos Dorf, Switzerland
\and
ETH Zürich, IPA, Hönggerberg campus, HIT J22.4, Wolfgang-Pauli-Str. 27,  8093 Zürich
\and ADNET Systems, Inc., Lanham MD, USA
\and Department of Mathematics, Physics and Electrical Engineering, Northumbria University, Ellison Place, Newcastle Upon Tyne, NE1 8ST, UK
\and
Department of Meteorology, University of Reading, Earley Gate, Reading, RG6 6BB, UK
\and
Donostia International Physics Center (DIPC), San Sebastián, Spain}

\date{Received XXX; accepted XXX}
 
  \abstract
    {}
   {We present an analysis of the first connection mosaic made by the SPICE instrument on board of the ESA / NASA Solar Orbiter mission on March 2$^{nd}$, 2022. The data will be used to map coronal composition that will be compared with \textit{in-situ} measurements taken by SWA/HIS  to establish the coronal origin of the solar wind plasma observed at Solar Orbiter. The SPICE spectral lines were chosen to have varying sensitivity to the First Ionization Potential (FIP) effect, and therefore the radiances of the spectral lines will vary significantly depending on whether the elemental composition is coronal or photospheric. We investigate the link between the behavior of sulfur with the hypothesis that Alfvén waves drive FIP \textcolor{black}{fractionation} above the chromosphere.} 
   {We perform temperature diagnostics using line ratios and Emission Measure (EM) loci, and compute relative FIP biases using three different approaches (two line ratio (2LR), ratios of linear combinations of spectral lines (LCR), and differential emission measure (DEM) inversion) in order to perform composition diagnostics in the corona. We then compare the SPICE composition analysis and EUI data of the potential solar wind sources regions to the SWA / HIS data products.}
   {Radiance maps are extracted from SPICE spectral data cubes, with values matching previous observations. We find isothermal plasma of around $LogT=5.8$ for the active region loops targeted, and that higher FIP-bias values are present at the footpoints of the coronal loops associated with two active regions. Comparing the results with the SWA/HIS data products encourages us to think that Solar Orbiter was connected to a source of slow solar wind during this observation campaign. \textcolor{black}{We demonstrate FIP fractionation in observations of the upper chromosphere and transition region}, emphasized by the behavior of the intermediate-FIP element sulfur.}
   {}


   \maketitle

\section{Introduction}\label{sec:intro}

The heliosphere is a low-density medium composed mainly of plasma, radiation, dust, and energetic particles. Characterizing this medium has been a long-standing goal in the field, with the aim of relating its properties to the magnetic activity of the Sun, as well as having a better understanding of the effects experienced by the various bodies immersed in this highly dynamic medium. The flows which generate the solar wind drive the evolution of the heliospheric magnetic field, and are therefore a prime source of information to answer the questions we are addressing in this paper.
The \textit{Solar Orbiter} (SolO) mission \citep{Muller2013, Muller2020, 2021A&A...646A.121G} with its six remote sensing and four \textit{in-situ} instruments provides a comprehensive diagnosis of solar plasma as well as solar wind plasma, providing insights into the origin of solar winds and the factors responsible for the generation and acceleration of the solar wind to speeds of hundreds of kilometers per second. Solar Orbiter will substantially build on the observations of the Helios mission \citep{WINKLER1976435}, taking the closest ever images of the Sun and directly observing the high latitude polar regions for the first time with its out of the ecliptic orbit.

As part of SolO's remote sensing suite, the SPectral Imaging of the Coronal Environment (SPICE) instrument \citep{SPICEpaper, Fludra_2021} is an imaging spectrometer that records extreme ultra-violet (EUV) spectra of the Sun. SPICE provides a complete temperature coverage, from the low chromosphere to the corona, as well as powerful off-limb coronal diagnostics. SPICE will produce Doppler velocity maps (e.g. \cite{1999Sci...283..810H}) as well as detailed composition diagnostics, thanks to its various simultaneously observable emission lines from low and high First Ionization Potential (FIP) elements. 

One of the strengths of the instrument is its ability to compare its data products with those from \textit{in-situ} instruments such as the Solar Wind  Analyser's Heavy Ion Sensor \citep[SWA--HIS;][]{Owen2020}. By linking remote sensing observations of the solar corona and transition region with \textit{in-situ} measurements of plasmas, fields, and energetic particles, it is possible to understand how solar processes are related to the plasma phenomena in the heliosphere and how the heliosphere is generated. 

Even though the origins of the fast solar wind are well known \citep{1973SoPh...29..505K, 2007ApJS..171..520C}, there are still open questions regarding the formation and mechanisms of the slow solar wind \citep{Abbo2016, 2017SSRv..212.1345C, 2020JASTP.20405271B}. SPICE's mosaic data will aid in explaining the formation of the slow solar wind, particularly by studying the FIP effect -- the enhancement of low-FIP elements in the corona \citep{2021SSRv..217...78P}, which has been a topic of intense investigation in recent years. 

In this paper, we present an analysis of the first composition mosaic from SPICE, and we demonstrate a variety of possibilities for future remote sensing observations of this nature. In Section 2, a description of the data and observations is provided, and we present and analyze the results in Section 3. Section 4 describes FIP-bias measurement techniques and results, \textcolor{black}{with a focus on the two active regions observed and the behavior of the element sulfur}. In Section 5, we discuss problems with the data, avenues of improvement, and future plans. We perform a partial replication of the analysis conducted in \cite{2022ApJ...940...66B} on November 2020 data (e.g. intensity diagnostics, EM loci analysis, and Mg/Ne measurements) and extend it by performing additional diagnostics on SPICE's new measurements, such as abundance and temperature diagnostics. In this paper, we seek to continue to build on previous work carried out on establishing new methodologies for interpreting measurements from SPICE.

\section{Data and observations}

SPICE operates in two modes: using a fixed slit position (\textit{sit-and-stare}); or by scanning the Sun in the direction perpendicular to the slit (\textit{rastering}) \citep{SPICEpaper, Fludra_2021}. It has four available slits of varying widths (2\arcsec, 4\arcsec, 6\arcsec, and 30\arcsec).

Two wavelength ranges are observed with SPICE: 704 to 790 \AA \, with the SW (short wavelength) detector, and 973 to 1049 \AA \; with the LW (long wavelength) detector. The spectral lines observable in both these ranges are produced by a number of different ions. These ions allow for the investigation of the relationships between all the observed layers of the solar atmosphere since they form at temperatures from 10\,000\,K (observable with SPICE's H I) to 10\,MK, reached by Fe XX (Table 1 of \cite{SPICEpaper}).

On March $2^{\text{nd}}$ 2022, SPICE performed three raster observations as part of the first occurrence of the Composition Mosaic SOOP (Solar Orbiter Observing Plans, \cite{2020A&A...642A...3Z}), which included all ten instruments: six remote sensing ones - SPICE, 
the Extreme UV Imager (EUI; \cite{2020A&A...642A...8R}),
the Polarimetric and Helioseismic Imager (PHI; \cite{Solanki_2020}),
the Spectrometer Telescope for Imaging X-rays (STIX; \cite{2020A&A...642A..15K}), 
Visible light and UV Coronagraph (Metis; \cite{2020A&A...642A..10A}) and the  
Heliospheric Imager (SoloHi; \cite{2020A&A...642A..13H}), and four in-situ ones:  
Solar Wind Analyser (SWA; \cite{2020A&A...642A..16O}), 
Radio and Plasma Wave analyser (RPW; \cite{2020A&A...642A..12M}),
Magnetometer (MAG; \cite{2020A&A...642A...9H}), 
Energetic Particle Detector (EPD; \cite{2020A&A...642A...7R}).

The composition mosaic observations lasted from 00:41 to 23:30 UTC while Solar Orbiter was very close to its first perihelion, at around a heliocentric distance of 0.54 AU. The data is available on the SPICE data release webpage\footnote{\url{https://doi.org/10.48326/idoc.medoc.spice.4.0}}.

SPICE has a spatial resolution along the slit of about 4\arcsec and a spectral resolution of 0.09 \AA. For this study, each recorded raster contains the same seven spectral intervals, or spectral windows, with an exposure time of 120 seconds. These spectral windows have been chosen to provide a uniform distribution of formation temperatures for the spectral lines, to sample multiple ions in each temperature range, and to provide the largest possible discrimination in FIP. Each window has a different spectral size, and in some of them a few spectral lines are blended. These blends require particular attention when computing the radiances of the spectral lines. A method to separate these blended lines will be presented in this section.

\begin{table*}[htbp]
\centering
\begin{tabular}{llllll}
\hline \hline
Spectral lines   & logT (K) & FIP (eV) & Transition   & Sub-windows ($\AA$)  & Spectral bin \\ \hline
\ion{O}{i} 988.6                 & 4.2      & 13.6     & 2s$^2$ 2p$^4$ $^3$P$_2$ - 2s$^2 2p^3$ $^2$D$_0$   & 987.4 -- 989.1       & (1,1)        \\
\ion{N}{iii} 989                 & 4.8      & 14.5     & $ 2s^2 2p ^2P_{1/2} - 2s 2p^2 \; ^2D_{3/2}$       & 989.1 -- 990.6       & (1,1)        \\
\textbf{\ion{N}{iii} 991} *      & 4.8      & 14.5     & $ 2s^2 2p ^2P_{3/2} - 2s 2p^2 \; ^2D_{3/2}$       & 990.6 -- 993.4       & (1,1)        \\
\ion{O}{iii} 702.8 *             & 4.9      & 13.6     & 2s$^2$ 2p$^2$ $^3$P$_{1}$ - 2s 2p$^3$ $^3$P$_{0}$ & 700.3 -- 703.1       & (1,1)        \\
\ion{O}{iii} 703.8               & 4.9      & 13.6     & 2s$^2$ 2p$^2$ $^3$P$_{2}$ - 2s 2p$^3$ $^3$P$_{1}$ & 703.1 -- 704.85      & (1,1)        \\
\ion{S}{iv} 748                  & 5.0      & 10.4     & 3s$^2$ 3p $^2$P$_{1/2}$ - 3s 3p$^2$ $^2$P$_{1/2}$ & 746.1 -- 749.0       & (1,1)        \\
\textbf{\ion{S}{iv} 750}         & 5.0      & 10.4     & 3s$^2$ 3p $^2$P$_{3/2}$ - 3s 3p$^2$ $^2$P$_{3/2}$ & 749.0 -- 752.3       & (1,1)        \\
\textbf{\ion{N}{iv} 765  $_1$} * & 5.1      & 14.5     & 2s$^2$ $^1$S$_{0}$ - 2s 2p $^1$P$_{1}$            & 762.37 -- 767.15     & (2,1)        \\
\textbf{\ion{S}{v} 786} *        & 5.2      & 10.4     & 3s$^2$ $^1$S$_{0}$ - 3s 3p $^1$P$_{1}$            & 784.2 -- 787.0       & (1,1)        \\
\ion{O}{iv} 787 *                & 5.2      & 13.6     & 2s$^2$ 2p $^2$P$_{1/2}$ - 2s 2p$^2$ $^2$D$_{3/2}$ & 787.0 -- 789.8       & (1,1)        \\
\ion{O}{vi} 1032 * $_1$          & 5.4      & 13.6     & 1s$^2$ 2s $^2$S$_{1/2}$ - 1s$^2$ 2p $^2$P$_{3/2}$ & 1029.14  --  1034.91 & (2,1)        \\
\ion{Ne}{viii} 770 *    & 5.8      & 21.6     & 1s$^2$ 2s $^2$S$_{1/2}$ - 1s$^2$ 2p $^2$P$_{3/2}$ & 767.2 -- 771.5       & (2,1)        \\
\ion{Mg}{viii} 772      & 5.9      & 7.6      & 2s$^2$ 2p \;$^2$P$_{3/2}$ - 2s2p$^2$\;$^4$P$_{5/2}$   & 771.5 -- 775.2       & (2,1)        \\
\ion{Mg}{ix} 706 *      & 6.0      & 7.6      & 2s$^2$ $^1$S$_{0}$ - 2s 2p $^3$P$_{1}$            & 704.85 -- 708.1      & (1,1)        \\ \hline \hline
\end{tabular}%
  \caption{Lines extracted from the mosaic data set, and the wavelength window used to determine the Gaussian parameters for the fitting. }
  \tablefoot{In bold are the four lines we use for the LCR relative FIP bias measurements. Lines marked with ‘*’ were used to infer the DEM. Lines with an index $_1$ indicate those which only require a single Gaussian fit, therefore the ones that should give the best results. Windows marked with a bin of (2,1) were binned in the spectral direction by a factor of two on-board, the rest were not.}
\label{tab:compo_study}
\end{table*}

The SPICE data pipeline provides calibrated data, the spectra have been corrected for slit tilts, detector misalignments and other distortions as well as flat-field and dark current \citep{SPICEpaper}.
\textcolor{black}{The pipeline includes a correction for the decline of the instrument's responsivity since the beginning of the mission.} This decline is happening uniformly across the SW and LW wavelength bands, affecting all lines equally. Consequently, the signal levels for a specific date close to the launch (2020) will differ from those in March 2022. We expect the disparity between November 2020 and March 2022 to be approximately a 2.5 factor. The burn-in (the degradation of the detectors due to long exposure to high intensities) is also to be taken into account, affecting mainly the N IV and Ne VIII lines, and to a lesser extent the N III and O III lines. Quantifying the burn-in factor of the last two lines poses a greater challenge due to the complexity of the region and the camera artifacts.

While looking at the rasters, horizontal-line noise prevented the data from being correctly processed. We partly addressed this issue by subtracting, from each y-$\lambda$ plane, the plane's minimum value; this removed some horizontal banding artifacts. We also binned down the rasters by a factor of two along the X-axis in order to improve computation times and signal-to-noise ratios. 

\subsection{Solar features observed in the mosaic}

During this first SOOP Mosaic, two active regions (ARs) were in SPICE's Field of View (FOV). NOAA 12958, on the "A" panel of the radiance maps depicted in Fig.~\ref{fig:intensity}, was caught during the first raster scan, and NOAA 12957 lies in between panel "B" and "C", \textcolor{black}{and lies close to a coronal hole, distinctly observable in the 193 \AA\;band of SDO/AIA}. These reported ARs were identified using the NOAA Active Region (NAR) database and the Solar Monitor software \citep{Gallagher2002}, and an extensive study on decayless kink oscillations by \cite{berghmans_first_2023} using EUI data has been conducted on those two ARs. Both ARs are simple bipoles, with a distinct separation between the positive and negative polarities, and hence are classified as $\beta$ regions. Their characteristics are summarized in Table \ref{CharacAR}.

\begin{figure}
\begin{minipage}[]{.45\linewidth}
\centering \includegraphics[width=4cm]{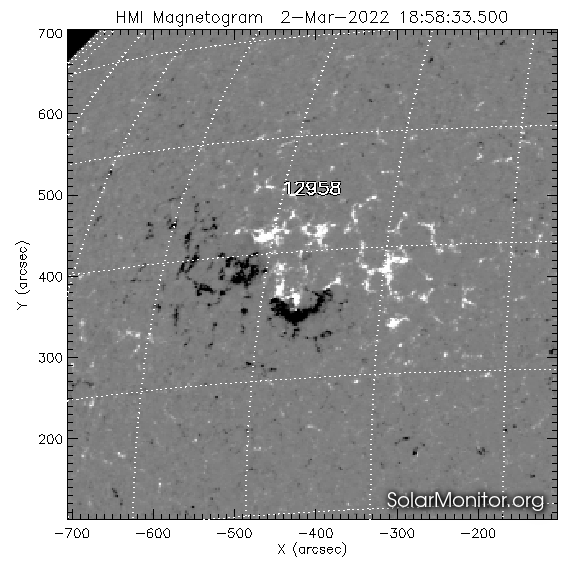}
\subcaption{AR 12958}
\end{minipage}\hfill  
\begin{minipage}[]{.45\linewidth}
    \centering \includegraphics[width=4cm]{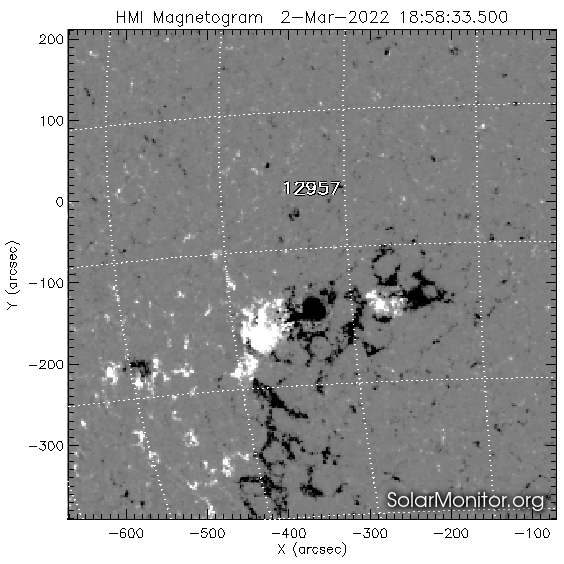}
        \subcaption{AR 12957}
\end{minipage}\hfill
\begin{minipage}[]{.45\linewidth}
\centering \includegraphics[width=4cm]{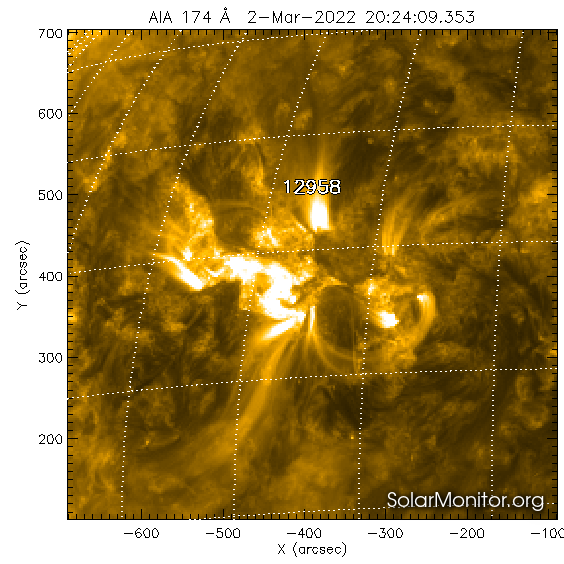}
\end{minipage}\hfill    
\begin{minipage}[]{.45\linewidth}
    \centering \includegraphics[width=4cm]{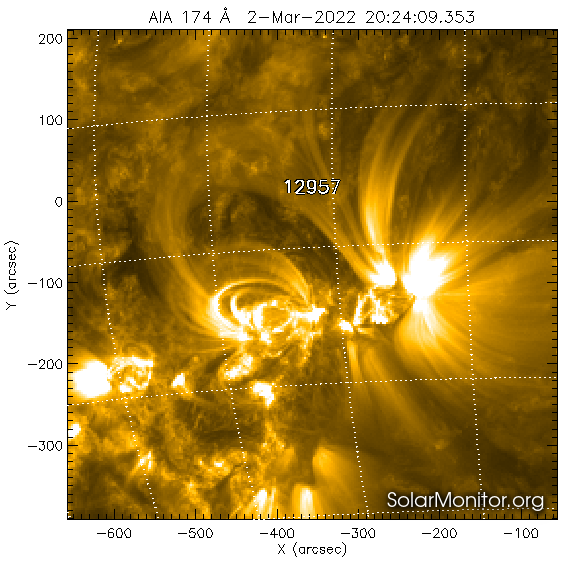}
\end{minipage}\hfill
\caption{Active Regions 12958 (a) and 12957 (b) observed on March 2$^{nd}$, \textcolor{black}{2022 at 20:24 UT by SDO/AIA imager at 171 \AA \, (representing the corona at 0.6MK), lower panel} and corresponding SDO/HMI images observed on March 2$^{nd}$, 2022 at 18:58 UT, \textcolor{black}{upper panel}. White/black areas are of positive/negative polarity.} 
\label{fig:HMIar}
\end{figure}

\begin{table}[htbp]
    \begin{tabular}{lll}
    \hline \hline
        Parameter & NOAA 12958 & NOAA 12957\\ \hline
        Concept &  Active Region & Active Region\\
        Hale class & $\beta$ & $\beta$ \\
        McIntosh class & Cro & Cro \\
        Solar X & -390\arcsec  & -356\arcsec  \\
        Solar Y & 403\arcsec  & -89\arcsec  \\
        Sunspot area & 20 & 30 (millionths)\\
        Number of spots & 2 & 5 \\
        Associated flares  & M2.0(17:31) & - \\
        \hline \hline
    \end{tabular}
    \caption{Technical description of observed ARs.} \label{CharacAR}
    \label{tab:class_AR}
\end{table}

The simple nature of the bipoles is clearly evident in the photospheric magnetic field data, in the top panel of Figure \ref{fig:HMIar}. We show the line-of-sight magnetogram taken by SDO/HMI \textcolor{black}{(Solar Dynmaics Observatory / Helioseismic and Magnetic Imager)} on March 2$^{nd}$ 2022 at 18:58 UT. From the magnetograms, it is evident that AR 12958 is a decaying field region whereas, AR12957 consists of multiple emerging structures. On AR 12957, two major bipoles can clearly be identified on the top panel of Figure 1 (b) and related to the loops that can be seen on the bottom panel of Figure 1 (b). Negative polarities (denoted in black) on the HMI images are also spatially consistent with the loops at the edges of the active region, which could be associated with open magnetic field.

\subsection{Spectral line fitting}
Once the SPICE pre-processing has been applied, a Gaussian fit of the emission peaks is performed. Two cases can be encountered, namely single or multiple line fitting. When a line has a narrow spectral size (typically under 25 pixels) and a single ion of interest is present, a single Gaussian is fitted onto the spectrum. If the window is wider but only one line is worth observing, a windowing of the spectrum is applied to define a particular region for the fitting. In the second case, several ions of interest are blended within a single window; we then fit the spectra with a model including several Gaussians. Regardless of the case, the continuum is modeled by a constant.

\begin{figure*}
    \centering 
    \includegraphics[width = 18cm]{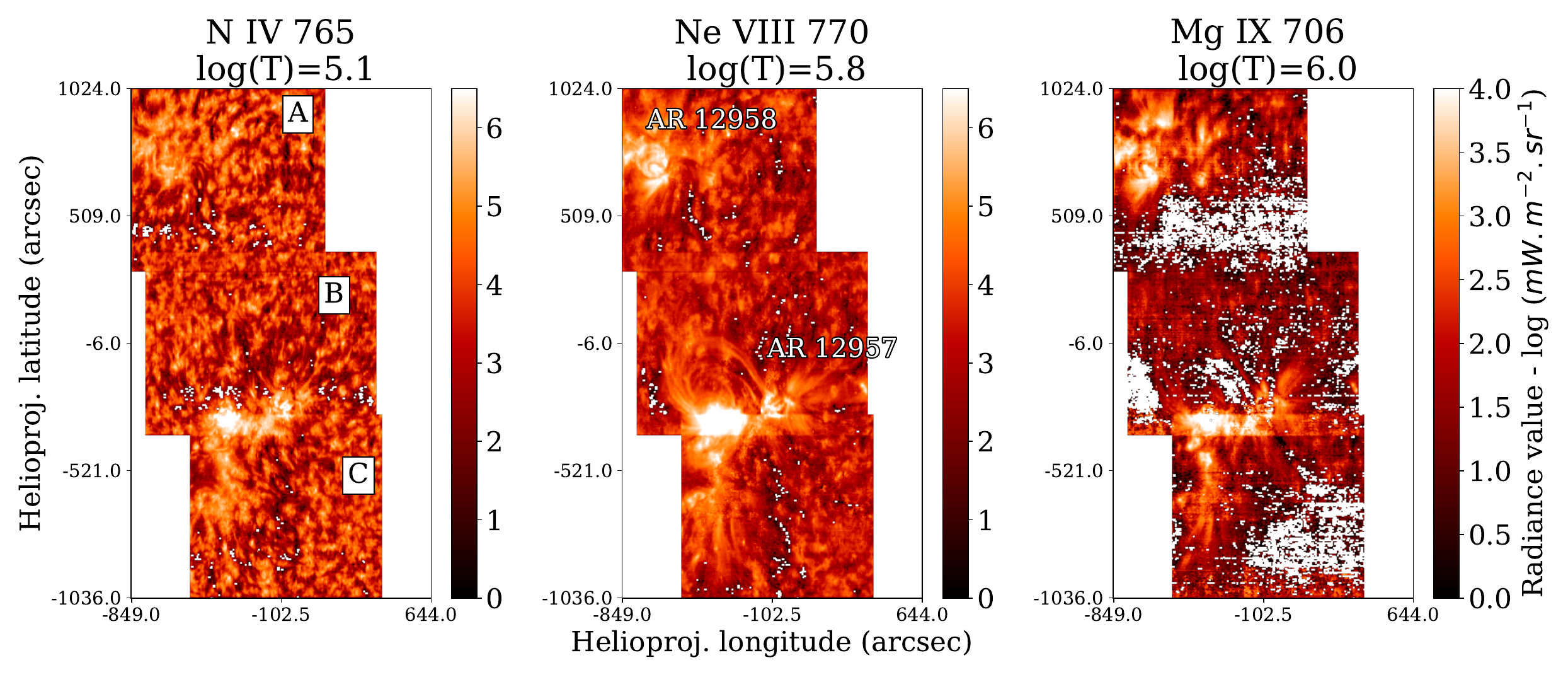}
    \hspace*{-1cm}
    \includegraphics[width = 18cm]{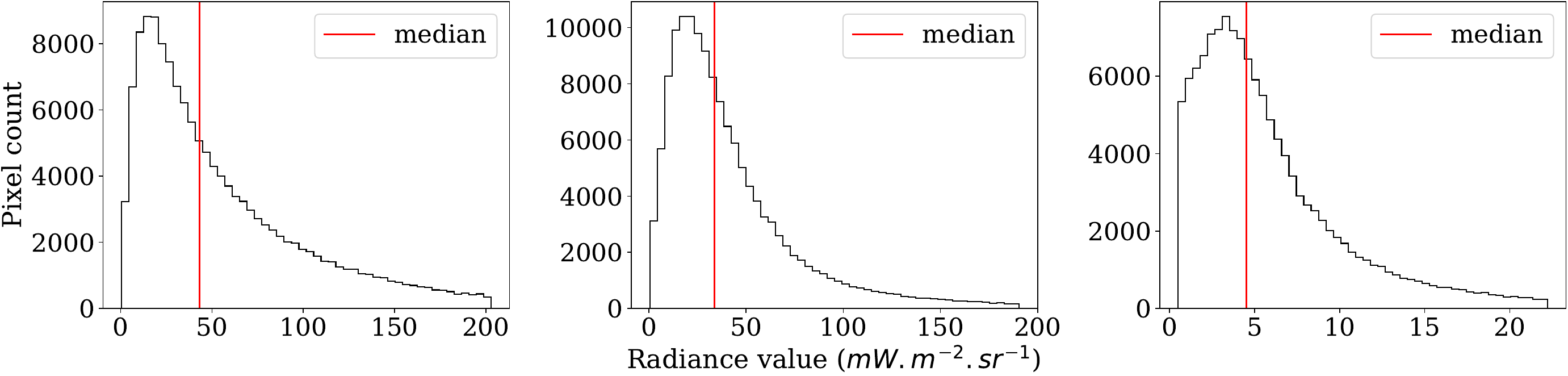}
    \caption{Top panel: radiance maps from lines N IV 765 \AA, Ne VIII 770 \AA \; and Mg IX 706 \AA. The three rasters are labeled A, B and C for more clarity. Bottom panel: histograms of corresponding radiances.}
    \label{fig:intensity}
\end{figure*}

 More details about the line fitting method and the windows are given in Appendix \ref{AppendixA}, and the main characteristics of the lines are shown in Table \ref{tab:compo_study}. Examples of the radiances derived from the fitting and their distribution are presented in Figure \ref{fig:intensity}, and the complete radiance diagnostic containing all of the extracted lines is shown in Appendix \ref{fullrad}. The chromospheric network can be clearly identified at lower temperatures (0.06-0.3 MK), and brighter, hotter loops start to be visible around 0.3 MK, (e.g. O VI 1032) line. All the lines show very bright moss -- low-lying, reticulated EUV emission \textcolor{black}{at} the base of hot and high-pressure loops, discovered in \textit{TRACE EUV} Fe IX/X images \citep{1999mossDePontieu} -- emission at the footpoints of the loops. Noticeably darker pixels represent unsuccessful fitting, mainly due to noisy spectra. 
\begin{table*}[htpb]
\centering
\begin{tabular}{llllll}
\hline \hline
Line & SUMER 1997 QS & SPICE 2022 QS & SPICE 2020 AR & SPICE 2022 AR S1 & SPICE 2022 AR S4 \\ \hline
        \ion{O}{iii} 702.8 & 27.5 & 13.5 $\pm$ 5.3 & 22.3 $\pm$ 6.6 & 31.0 $\pm$ 7.7 & 139.7 $\pm$ 17.1 \\ 
        \ion{O}{iii}  703.8 & 43.5 & 16.4 $\pm$ 5.2 & 22.7 $\pm$ 6.5 & 38.5 $\pm$ 7.6 & 132.6 $\pm$ 12.9 \\ 
        \ion{Mg}{ix} 706 & - & 3.6  $\pm$ 5.3 & 19.1 $\pm$ 5.1 & 30.2 $\pm$ 7.7 & 33.9 $\pm$ 14.4 \\ 
        \ion{S}{iv}  748 & 2.9 & 3.8  $\pm$  1.4 & - & 13.3 $\pm$ 7.3 & 10.9 $\pm$ 14.8 \\ 
        \ion{S}{iv}  750 & 6.9 & 5.2  $\pm$  1.9 & - & 17.7 $\pm$ 7.4 & 46.1 $\pm$ 14.2 \\ 
        \ion{N}{iv} 765 & 80.7 & 29.6  $\pm$  4.3 & 32.8 $\pm$ 8.6 & 123.7 $\pm$ 9.7 & 394.9 $\pm$ 16.1 \\ 
        \ion{Ne}{viii} 770 & 73.1 & 30.3 $\pm$ 7.3 & 67.2 $\pm$ 17.2 & 361.7 $\pm$ 11.5 & 1087.4 $\pm$ 31.1 \\ 
        \ion{Mg}{viii} 772 & - & 9.3 $\pm$ 1.7 & 10.4 $\pm$ 3.3 & 57.2 $\pm$ 12.0 & 110.0 $\pm$ 29.0 \\ 
        \ion{S}{v} 786 & 32.1 & 14.3 $\pm$ 9.9 & 21.7 $\pm$ 6.2 & 77.5 $\pm$ 16.4 & 81.4 $\pm$ 87.5 \\ 
        \ion{O}{iv} 787 & 58.5 & 26.1 $\pm$ 9.9 & 38.6 $\pm$ 10.3 & 121.1 $\pm$ 16.4 & 187.6 $\pm$ 86.9 \\ 
        \ion{O}{i} / \ion{Na}{vi} 988.6 & - & 6.3 $\pm$ 14.8 & - & 35.3 $\pm$ 22.9 & 45.4 $\pm$ 43.3 \\ 
        \ion{N}{iii} 989.8 & 23.0 & 10.4 $\pm$ 14.5 & - & 40.4 $\pm$ 22.2 & 175.5 $\pm$ 40.2 \\ 
        \ion{N}{iii} 991.5 & 44.2 & 23.0 $\pm$ 14.6 & 12.3 $\pm$ 4.8 & 120.5 $\pm$ 22.4 & 472.3 $\pm$ 77.4 \\ 
        \ion{O}{vi} 1032 & 354.0 & 73.3 $\pm$ 11.2 & 273.7 $\pm$ 69.0 & 1562.0 $\pm$ 33.1 & 3046.5 $\pm$ 61.3 \\ \hline \hline
\end{tabular}
\caption{Comparison of radiance values (see Figure \ref{fig:S1S2} for the regions selected, SPICE QS including both S2 and S3).} 
\tablefoot{The radiances are compared with the SUMER instrument ones (QS), taken from \cite{dufresne2023benchmark} and SPICE 2020 (AR) from the region S2 of \cite{2022ApJ...940...66B}. Units of radiances for SUMER are in erg sr$^{-1}$ cm$^{-2}$ s$^{-1}$ and SPICE ones have been converted to mW sr$^{-1}$ m$^{-2}$, which are equivalent. Our values are corrected by a degradation factor of 2.5 and the spectral binning of the lines N IV, Ne VIII / Mg VIII and O VI has been accounted for by multiplying the radiances by the binning factor.}
\label{tab:comp_rad}
\end{table*}

\begin{figure}
\hspace*{-0.5cm}
    \centering
    \includegraphics[width=\linewidth]{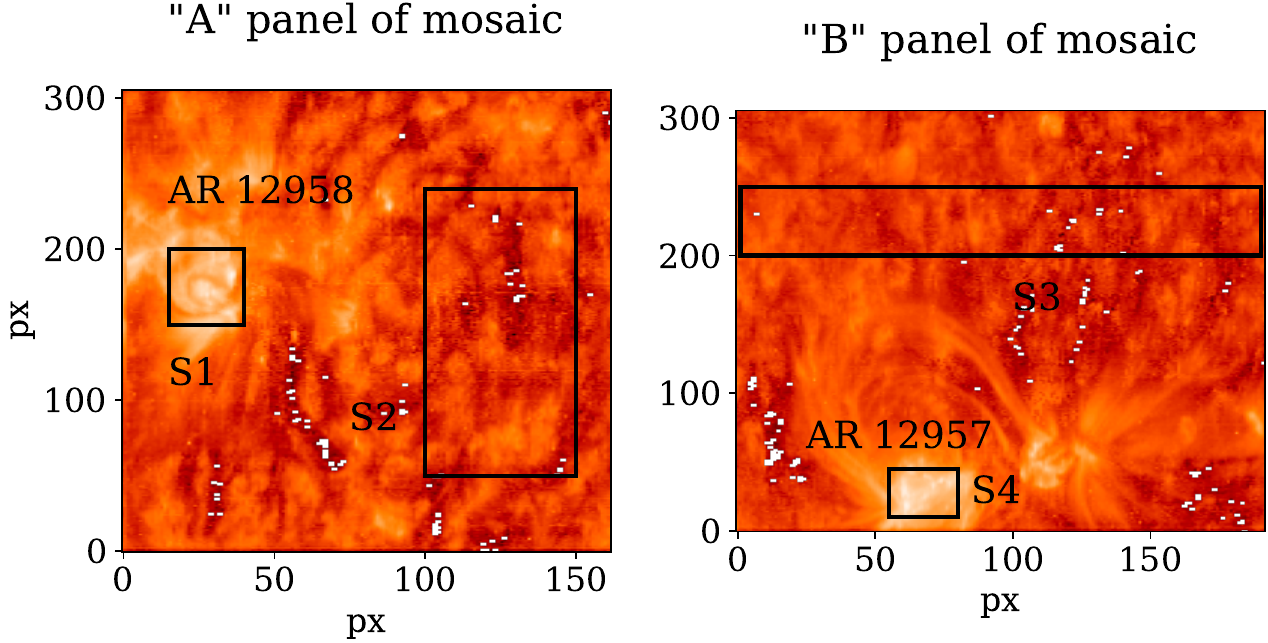}
    \caption{Areas S1 (AR), S2 / S3 (QS) and S4 (AR), seen in the Ne VIII 770 \AA\;line.}
    \label{fig:S1S2}
\end{figure}

To validate the results of the line fitting, we compare our values with previous SPICE measurements from \cite{2022ApJ...940...66B} and the compilation of quiet Sun radiances, mostly from SUMER, listed by \cite{dufresne2023benchmark}. To compare the values of the area \textit{S1} observed in \cite{2022ApJ...940...66B} (AR 12781), we took two similar active region areas, denoted S1 and S4 on Figure \ref{fig:S1S2}. SUMER values are computed over a quiet Sun area; to match these measurements, two areas from panels A and B (S2 and S3, see Figure \ref{fig:S1S2}) have been selected, and the average of the intensity values have been computed. The results are presented in Table \ref{tab:comp_rad}. Even though we tried to choose similar regions to average the radiance values, it is important to keep in mind that the Sun is an active star and that we cannot expect to obtain the exact same values, especially taking into account the current stage in the solar cycle. The deviation of the values is significantly greater for the active regions. The sizes of the ARs observed are fairly similar, but their magnetic structures are quite different. The radiance values of the cooler lines overlap within the uncertainties, but brighter, hotter lines show factors of two, up to 5, which is not uncommon when observing ARs.

The quiet Sun values should show less discrepancy than the ARs, especially for lower temperature lines \citep{2014A&A...563A..26A}. \textcolor{black}{The SUMER values have been taken during February 1997, at solar minimum, whereas the 2022 data corresponds to rising activity. Hence, the difference in their solar cycle stage could be an explanation for the discrepancies observed, which we are currently unable to explain.} One cause of these differences could also be that the dark subtraction for SPICE was not very accurate for this dataset, resulting in inaccuracies in the intensities, especially for quiet Sun values, which have a lower signal. \textcolor{black}{These issues are further discussed in section \ref{discussion}}. Another possibility is that the quiet Sun region selected here is not representative of the average quiet Sun. Further observations with SPICE might resolve this issue.\\

In summary, after applying some pre-processing to the spectra, we performed a Gaussian fitting of the emission peaks. The results of the line fitting are compared with previous SPICE and SUMER measurements, showing acceptable coherence on the relative intensity between lines for the ARs, but differences between the quiet Sun values which should be further investigated.

In the following sections, we present the diagnostic methods used to infer plasma properties from SPICE mosaic data.

\section{Methods}

The intensity of an optically thin spectral line, assuming a plasma in ionization equilibrium, and the abundance of an element $X$ to be constant along the line of sight, can be defined as:
\begin{equation}\label{eqInt}
\begin{split}
    I(\lambda_{ji}) & =  \mbox{Ab}(X)\,\int_h  C(n_e, T, \lambda_{ji}) \, n_e \, n_H \, dh \\
                    & \approx \mbox{Ab}(X)\,\int_h  C(n_e, T, \lambda_{ji}) \, n_e^2 \, dh \,, 
\end{split}
\end{equation} 
$C$ being the contribution function, $\mbox{Ab}(X)$ the elemental abundance of the element $X$ and $n_e$ and $n_H$ the electron and hydrogen density, respectively. We used the \cite{2015Chianti} CHIANTI database for every calculation including the contribution function and / or the abundances. An estimation of the plasma temperature, assuming an isothermal distribution of the plasma, can be inferred from the ratio of two lines of the same element, as their ratio is directly equal to the ratio of their contribution function:
\begin{equation}\label{eqtemp_ratio}
    \frac{I^{obs}_1}{I^{obs}_2} = \frac{C_1(T_0, n_e)}{C_2(T_0, n_e)}
\end{equation}

As defined in \cite{del_zanna_solar_2018}, a column differential emission measure (DEM) can be defined if a relationship exists between the temperature and the electron density:
\begin{equation}
    \mbox{DEM}(T)   = n_e^2 \, \frac{dh}{dT} \,.
\end{equation}

Equation \ref{eqInt} can then be rewritten as:
\begin{equation}
    I(\lambda_{ji}) = \mbox{Ab}(X) \int_T  C(n_e, T, \lambda_{ji})\,\mbox{DEM}(T) \, dT.
\end{equation}

A column emission measure can also be computed, integrating the DEM over the temperature range:
\begin{equation}
    \mbox{EM} = \int_h n_e^2 \, dh = \int_T \mbox{DEM}(T) \, dT.
\end{equation}

The DEM represents the amount of plasma along the line of sight, making it an essential parameter for inferring diagnostics from observed radiances. 

A simple estimation of the temperature distribution  (first introduced by \cite{strong_1980loci}) is to plot the ratio $I_{obs} / G(T)$ as a function of temperature, with $G(T) = Ab(X)\,C(T)$. The loci of the resulting curves represent an upper limit to the emission measure at each temperature.
In an isothermal plasma, the curves should intersect at a common point. If the plasma is not isothermal, the curves are more uniformly distributed; it is then necessary to perform a DEM analysis to obtain the distribution of plasma along the line of sight.

The EM loci method offers a simple way to measure the relative abundances of elements.
It is also a far more accurate method than any other one, when the plasma is isothermal. In fact, incorrect results have been published in the literature when forcing 
a continuous (DEM) distribution to solar features such as 
active region cool loops, as reviewed in \cite{del_zanna_solar_2018}.

\begin{figure*}[htpb]
\hspace*{-1cm} 
    \centering
    \includegraphics[width=18cm]{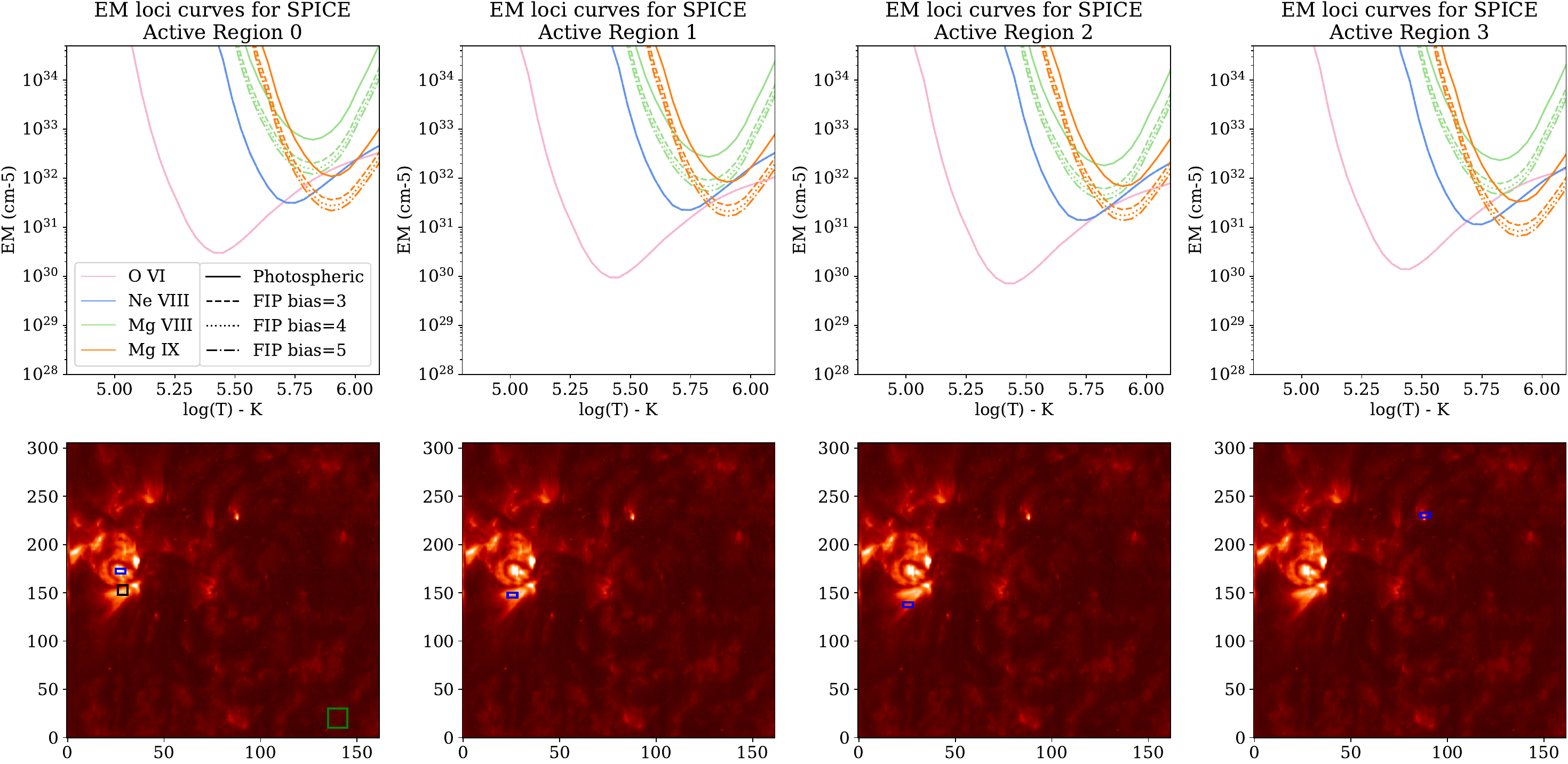}
    \caption{EM loci curves for high temperature lines of SPICE, \textcolor{black}{over AR 12958. Blue rectangles on the maps denote the area over which the curves are computed. Curves of low-FIP elements are plotted with different FIP bias values (1 (photospheric), 2, 4 and 5)}. The intersection of the curves (especially for a FIP bias of 4) indicates an isothermal plasma around logT=5.8. The black and green boxes on the bottom left panel are respectively the areas chosen to compute the DEM for AR and QS region depicted in Figure \ref{fig:DEMinAR}.}
    \label{fig:DEMloci}
\end{figure*}

The EM loci curves for AR areas from the connection mosaic are shown in Figure \ref{fig:DEMloci}. The computation has been performed using photospheric abundances for high-FIP elements (elements having high first ionization energy, \textcolor{black}{typically above 10 eV)}, and different FIP bias values for the low-FIP (elements having low first ionization energy, \textcolor{black}{below 10 eV}) ones -- usually, a factor of 4 is adopted \citep{2016ApJ...824...56W}.

Caution should be taken when considering the Li-like ions (Ne VIII and O VI) as they have often been found anomalous in terms of their emission measures, as reviewed in \cite{del_zanna_solar_2018}. 
New ionization equilibria presented by 
 \cite{dufresne2023benchmark} improve the results, but still leave significant discrepancies between the observed and predicted intensities for some Li-like ions. 
 The reasons are still being investigated, but could be linked to the high-temperature tail of their contribution function or time-dependent ionization. 

For active region loops, the curves overlap at a unique temperature value, which is an indication of a near isothermal plasma in the area observed, as expected \citep[see the][review]{del_zanna_solar_2018}. The values and shape of the curves are consistent with the ones found in \cite{2022ApJ...940...66B}, especially for the FIP bias values between 3 and 4, in agreement with \cite{2016ApJ...824...56W} and \cite{Baker_2015}.
The intersection of the curves indicates a plasma temperature slightly lower than 1 MK, around log $T$ = 5.8 MK.

While quantifying the DEM should offer much more reliable information than other techniques (e.g. single line ratios), such as more accurate elemental abundances, unfortunately the DEM estimation and inversion is a complex problem; it is very difficult to estimate the DEM properly, especially when using an automated method. Issues have already been investigated and discussed by \cite{1997ApJ...475..275J}, \cite{2002A&A...385..968D}, \cite{Guennou_Auchère_Soubrié_Bocchialini_Parenti_2012}, \cite{Guennou_Auchère_Soubrié_Bocchialini_Parenti_Barbey_2012} and \cite{del_zanna_solar_2018}. 
We were also unable to de-blend some of the SPICE spectral lines used to infer the DEM, and thus we might be overestimating their radiances. Several additional lines were included for the computation of the contribution functions in order to take into account the blends present in our set of lines.

In addition to that, uncertainties, both random and systematic, and limited temperature resolution make the inference of the DEM even more challenging \citep{1997ApJ...475..275J, 1991A&A...249..277B}.

This problem therefore needs strong constraints, which are set in our case by using the strongest available lines of low and high FIP: O III, Mg IX, N IV, Ne VIII, Mg VIII, S V, O IV and N III. This set of lines gives us good constraints at lower temperatures but lacks some high-temperature ones. In further work, using the HRI 174 \AA\;band from EUI could provide a good, reliable high-temperature constraint for the DEM. 

For this work, we used a recent method for DEM reconstruction \citep{Plowman_2020}, which showed satisfying results on SDO/AIA data. The temperature range used for the DEM inversion is $10^{4.7}$ to $10^{6.2}$ K with a temperature bin size of $\Delta(\log T) = 0.035$ K.  
As a simple test of the accuracy of our results, we can compare the usual DEM shape within an active region to the one we compute over AR 12958. The DEMs that we computed are intended to give an indication of which lines contribute the most to SPICE's spectra under different solar conditions (CH, QS, AR).

\begin{figure}[H]
    \includegraphics[width=0.9\linewidth]{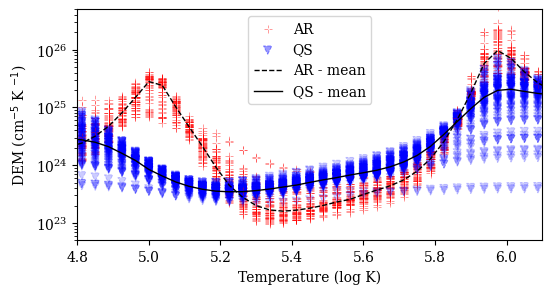}
    \caption{Distribution of the DEM with temperature over a QS and AR area. The QS curve has a flatter, more multithermal-like shape, whereas the AR curve has two clear emission peaks. For regions selected, see Figure \ref{fig:DEMloci}.}
    \label{fig:DEMinAR}
\end{figure}
The distribution of DEM values according to temperature in this active region is shown in Figure \ref{fig:DEMinAR}.

The peak (bell-shaped) coming from the AR emitting at logT = 6.0 is what we would expect for our temperature range and the lines chosen -- a higher temperature is observed over the active region targeted. The DEMs over a QS area present a large local minimum above logT = 5.2 and a peak above logT = 6.0 as seen in \citep{1981ApJ...245.1141R}.
We also observe an increase as it reaches chromospheric temperatures, as described in \cite{del_zanna_solar_2018}. Caution should be taken when treating the low-temperature tail of the DEM distribution, as it is directly constrained by low-temperature lines that may be influenced by opacity effects. Another issue is that the two strongest lines used in the DEM inversion are Li-like ions, which have a significant contribution at temperatures higher than the peak ones, due to the ionization fraction tail resulting from dielectronic recombination. Anomalies are discussed in \cite{2002A&A...385..968D} and references therein.

In the next section, we conduct further analysis based on the concepts presented previously, involving solar active regions description, temperature diagnostics, and comparison with EUI data and future \textit{in-situ} HIS data.

\section{Preliminary diagnostics and comparison with other instruments}

\subsection{Comparison with EUI data}
On the very same day, the Extreme UV Imager (EUI)  instrument recorded synoptic and high-resolution images. EUI has three telescopes: the Full Sun Imager (FSI), which records images at two wavelengths (174 and 304 \AA), and two high-resolution imagers. The FSI images were taken with a frequency of one image every 900 seconds.

\begin{figure*}
\centering
    \includegraphics[width=0.7\linewidth]{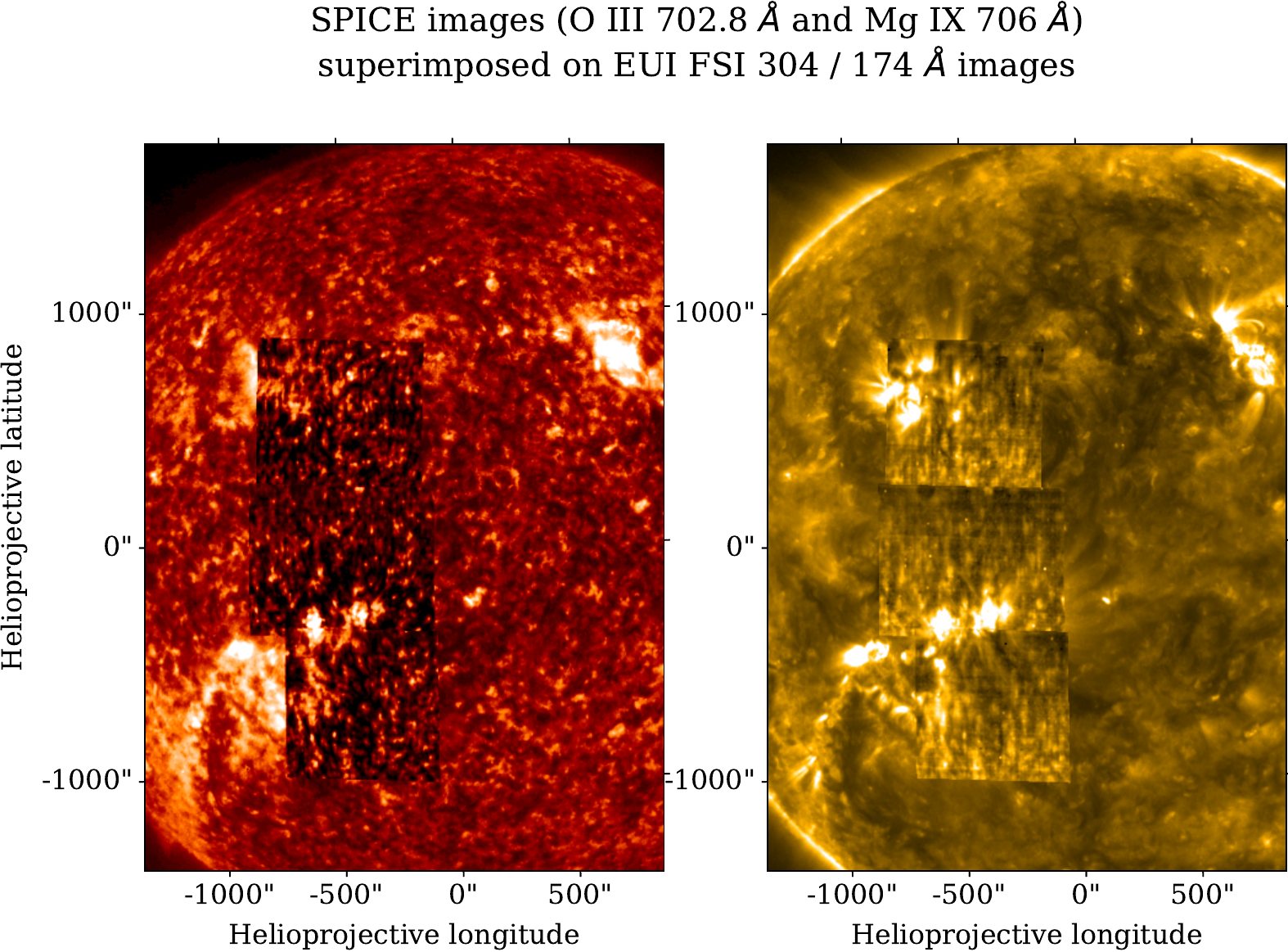}
    \caption{Left panel: SPICE O III 703 \AA\; superimposed on EUI FSI 304 \AA. Right panel: SPICE Mg IX 706 \AA\; on top of EUI FSI 174 \AA.}
    \label{fig:spiceEUI}
\end{figure*}

To check the accuracy of our line fitting results, we compare the SPICE data with the EUI on Solar Orbiter images. The higher resolution of EUI images allows for linking of precise structures to the composition measurements of SPICE.
However, EUI images are recorded almost instantly while SPICE's rasters are recorded over several hours, which causes an alignment problem between the two resulting images. To address this problem, we used the \texttt{reproject} package available in Python \citep{thomas_robitaille_2023_7584411}, and added to it an empirical ad-hoc offset to clear out any remaining misalignment. The results of this reprojection are presented in Figure \ref{fig:spiceEUI}. We selected approximately co-temporal FSI images, and the SPICE spectral lines plotted on top of EUI images have been chosen so that the temperature of the ions observed match. For the EUI 304 \AA\; which corresponds to He II (logT = 4.9), SPICE's O III 702.8 \AA\; (logT = 4.9) is superimposed. The targeted region on this left panel is the photosphere and the lower transition region. As for EUI 174 \AA\;(Fe X, logT = 6.1), SPICE's Mg IX 706 (logT = 6.0) is superimposed, despite its low SNR. In the latter, we look for upper transition region and coronal activity. The superpositions show very good consistency: we observe the same texture patterns on the QS areas and the active regions' shape and location match almost perfectly. We look forward to comparing future SPICE images with HRI images in order to get more detailed comparisons. These superpositions nonetheless require further processing (we applied a Contrast Limited Adaptive Histogram Equalization (CLAHE) on the normalized image, which performs local contrast enhancement using histograms computed over different tile regions of the image) to match the contrast of the two images; especially for the higher temperatures Fe X/Ne VIII which have narrower bands.

\subsection{Back-mapping with \textit{in-situ} data}

The comparison of SPICE abundance ratio maps with \textit{in-situ} HIS abundances offers a potential avenue for confirming the magnetic connectivity between the solar wind observed at a given instrument location and the wind sources on the Sun. \textit{In-situ} measurements are all the more important because once the solar wind plasma leaves the Sun (above 1.5 solar radii), abundances are not affected by any fractionation (or FIP effect) processes, opening an independent window for composition diagnostics \citep{2019BakerD, 2021SSRv..217...78P}. However, remote and direct measurements need to be connected since they do not record the same events at the same time. Previous studies involving data from the SWICS instrument on board of \textit{Ulysses} and ACE satellites could be used as reference points.

\begin{figure}
    \hspace*{-0.5cm} 
    \centering
    \includegraphics[width=0.999\linewidth]{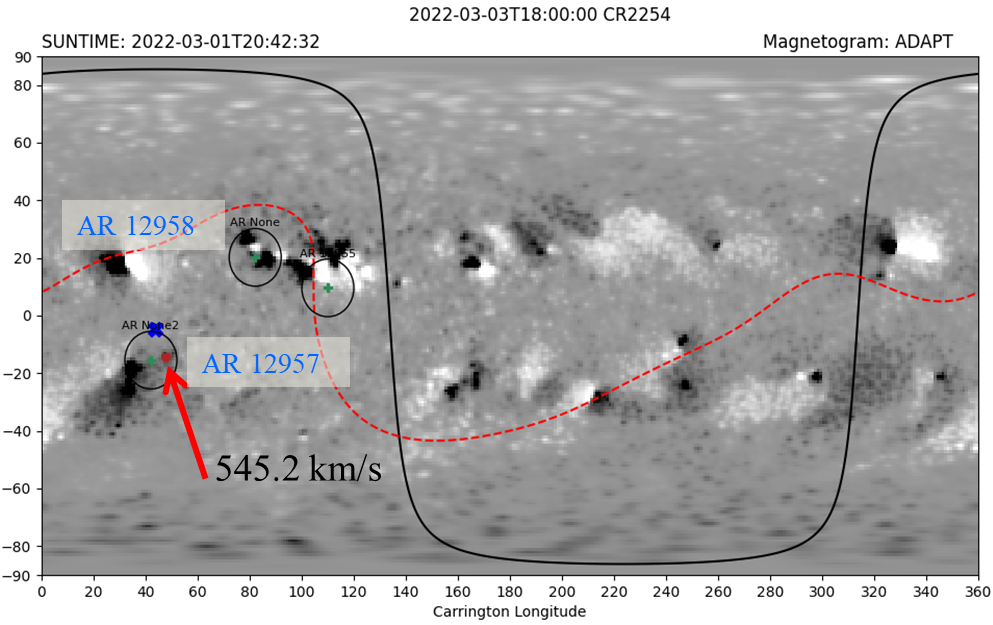}
    \hspace*{-0.5cm} 
    \centering
    \includegraphics[width=\linewidth]{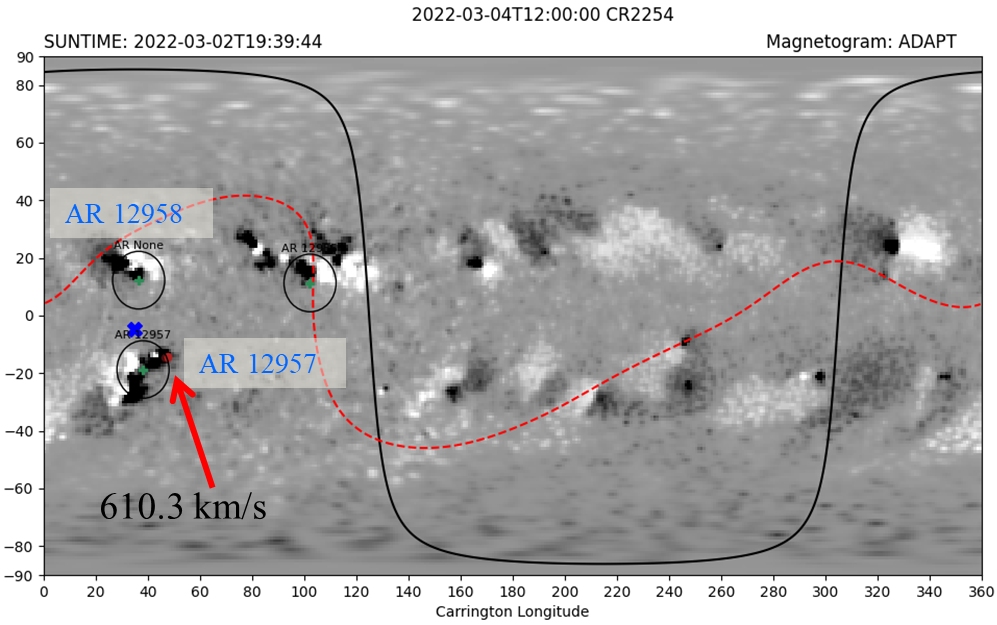}
    \hspace*{-0.5cm} 
    \centering
    \includegraphics[width=\linewidth]{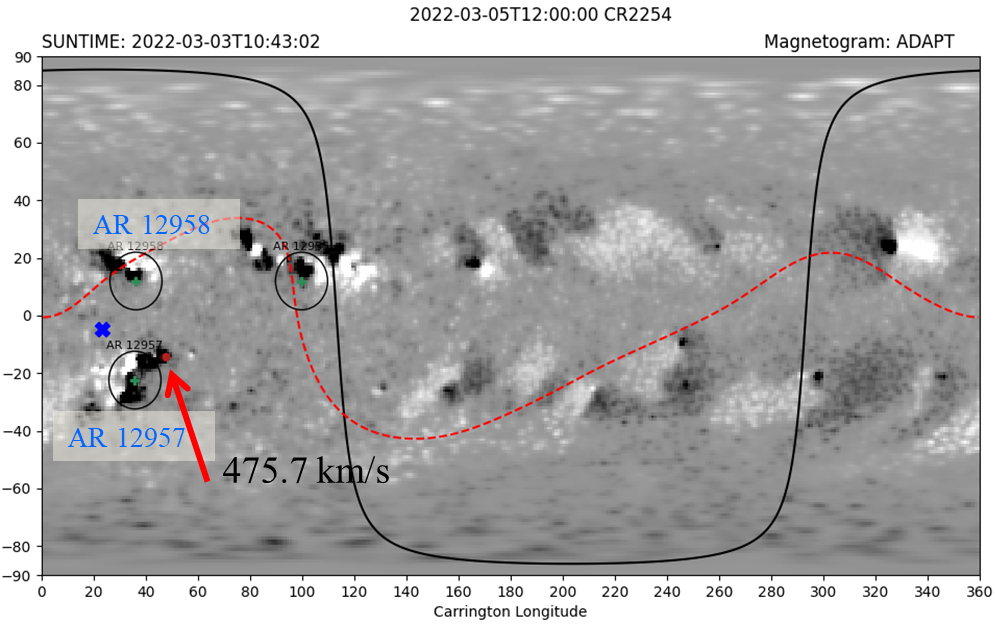}
    \caption{Connectivity maps from the MADAWG Connectivity Tool, from March 1$^{\text{st}}$, 2022 to March 3$^{\text{rd}}$, 2022. The maps are plotted over ADAPT magnetograms. The blue cross represents the sub-spacecraft point, and the red dot (pointed by the red arrow) the footpoint of the measured solar wind. The red dashed line represents the Heliospheric current sheet, and the black solid curve the visible disk from SolO.}
    \label{fig:adapt_map}
\end{figure}

In order to correctly back-map the events observed with remote sensing instruments, we need to take into account the delay between the two observations of the same event. To model the connectivity of Solar Orbiter, we used the MADAWG connectivity tool \citep{2020A&A...642A...2R}. The coronal magnetic field is reconstructed by the tool using the potential field source surface (PFSS) model and then extended assuming a Parker spiral model. The tool provided an estimated propagation time of two days and 6 hours between the Sun and Solar Orbiter. The evolution of AR 12957 and AR 12958 can be followed with the tool, as shown in Figure \ref{fig:adapt_map}, including the solar wind footpoint speed. The SW speed forecast increases starting at 545.2 km/s on March 1$^{\text{st}}$, reaching 610.3 km/s at the time of SPICE's observations, and then decreasing to 475.7 km/s on March 3$^{\text{rd}}$. SolO's connectivity point stays connected to the boundary of \textcolor{black}{the emerging }AR 12957, first to the negative polarity and moving on to the second negative polarity of the active region, adjacent to an equatorial coronal hole. The radial magnetic field measured by MAG supports this as it is predominantly negative throughout the time period.

\begin{figure*}
    \centering
    \includegraphics[width = \linewidth]{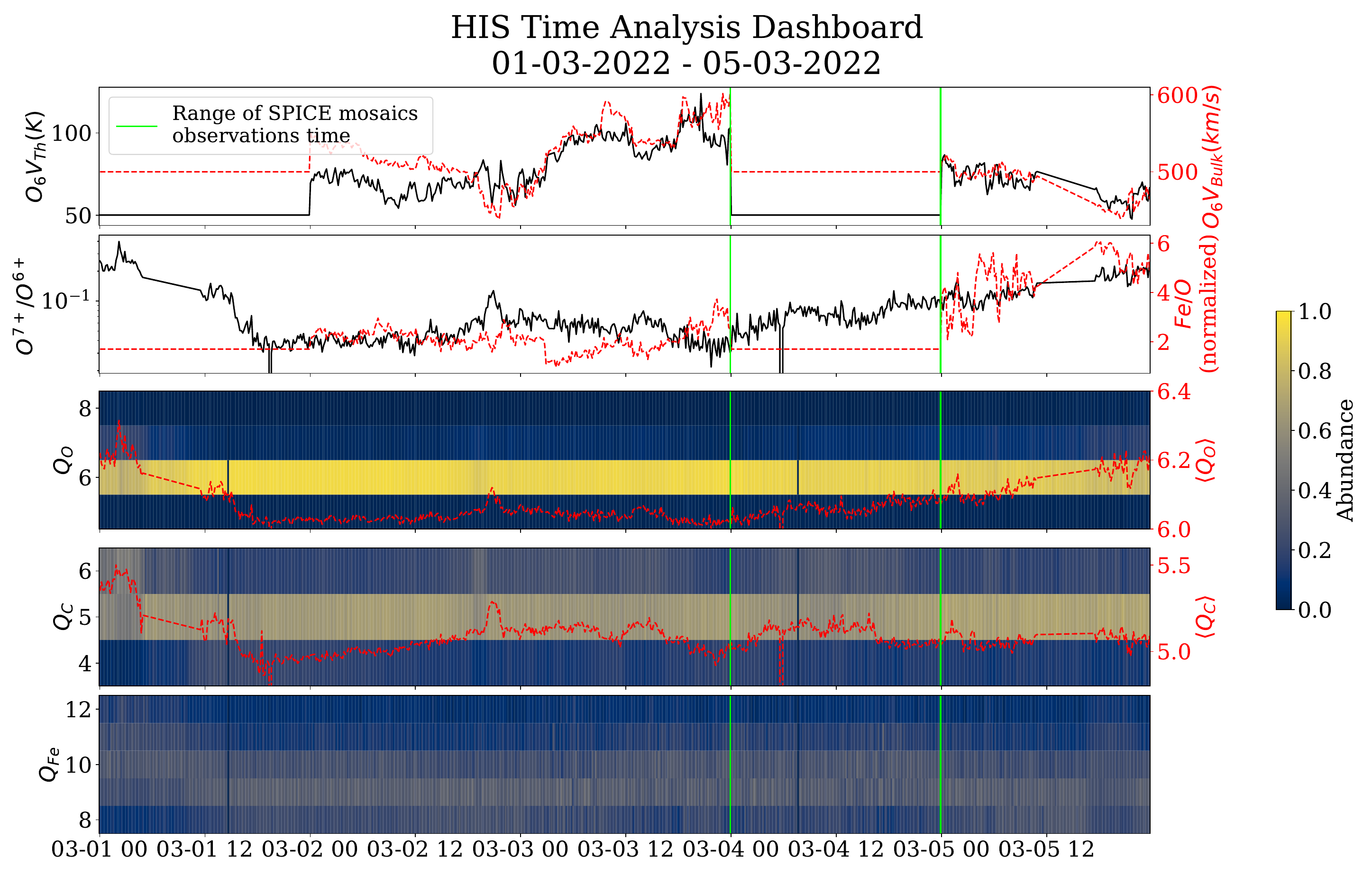}
    \caption{\textcolor{black}{Multi-panel plot from HIS data recorded from March 1st 00:00 UTC to March 6th 00:00 UTC. Top panel: $O^{6+}$ thermal and bulk velocity, second panel: $O^{7+}/ O^{6+}$ ratio and elemental Fe/O ratio (which has been normalized by the photospheric value 0.059 \citep{Asplund2021} for easy comparison with FIP bias values). The three bottom panels show the charge state distribution of elements O, C and Fe, as well as their average for O and C ($<Q_O>, <Q_C>$).}}
    \label{fig:HIS_panel}
\end{figure*}

The HIS and SPICE instruments have a wide range of joint observations, as HIS measurements can include He$^{1-2+}$, C$^{2-6+}$, N$^{2-7+}$, O$^{2-8+}$, Ne$^{6-9+}$, Mg$^{6-12+}$, Si$^{6-12+}$, S$^{6-14+}$, and Fe$^{3-24+}$. Consequently, when SPICE and SWA/HIS are employed together, the intensities recorded by SPICE can be used to validate the solar wind model predictions in the transition region and low corona. Meanwhile, HIS measurements can assess the predicted charge states in situ, as explained in \cite{2018IAUS..335...87F}. 
Unfortunately, SWA/HIS data corresponding to the dates of SPICE connection mosaic observations are largely incomplete. An example of an analysis panel of SWA/HIS data is shown in Figure \ref{fig:HIS_panel}. The data displayed show the evolution from March 1$^{st}$ 2022 to March 6$^{th}$ 2022. At the moment, we can only infer assumptions about the connection between the two instrument's data.

\textcolor{black}{Omitting the gaps, during SPICE observations (between the two green lines in Fig. \ref{fig:HIS_panel}), we can still see a substantial increase in the O$^{7+}$/O$^{6+}$ ratio, indicating a higher ionization of the plasma and a higher electron temperature. The Fe/O ratio, which could be considered as an approximation of the FIP bias, seems to show a growing trend that translates into an enhancement of low-FIP elements. As discussed in \cite{2019ApJ...879..124L}, an enhancement in the Fe/O ratio would be a good indicator for an open slow solar wind flux if the spacecraft is linked to an open-field region. Also, the O$^{6+}$ bulk velocity, which can be taken as a proxy for the outflow speed, is increasing starting March 3$^{\text{rd}}$ and then slowing down after March 4$^{\text{th}}$, corresponding to the top panel on Figure \ref{fig:adapt_map}. The increase in Fe/O and in the average charge states, the decrease in the flow velocity encourages us to think that Solar Orbiter shifted from a fast solar wind source region to a slow solar wind one, and this is what we will investigate more in detail in Section \ref{sectionFIP}.}

Looking further \textcolor{black}{ahead}, SPICE and HIS will offer the capability to determine the abundance ratios of low-FIP and high-FIP elements (e.g., Fe/O, S/N, Mg/Ne). By comparing the abundance ratio maps obtained by SPICE with the in situ abundances measured by SWA/HIS, it would be possible to verify the magnetic connectivity between the solar wind observed at the spacecraft's location and the solar sources of the wind. 

We also expect HIS data to provide elemental composition and ionic and proton temperature signatures for the active regions we observed. In particular, higher values of the ratio O$^{7+}$/O$^{6+}$ (indicating higher coronal temperatures), and higher ionized elements are expected, as well as distinct peaks in speed and density corresponding to the flares observed. To compare charge states between HIS and SPICE, it is necessary to consider their measurement altitudes and the freeze-in phenomenon \citep{2020A&A...642A...2R}. The HIS measurements capture charge states at altitudes above the freeze-in region, while SPICE measurements occur below this region. To bridge the gap, additional information from a coronal electron temperature model is needed. We also look forward to comparing our abundance ratios (especially Mg/Ne, O/Ne, the latter being dependent of the solar wind speed) with their HIS equivalents. However, we do not expect a massive change in high-FIP elements relative abundances, and only a subtle one for low-FIP elements. Going forward, we expect to have more complete HIS data sets to compare with future SPICE mosaic observations.

\subsection{Temperature diagnostics}

Temperature diagnostics can be obtained by using a ratio of two lines coming from the same element at two different ionization stages, if their contribution function does not depend on density, and assuming an isothermal plasma ($T = T_0$) along the line of sight. This assumption has to be taken with caution, as it is valid only for specific regions (e.g., active region loops).
The isothermal assumption being made, a temperature $T_0$ can directly be deduced from the ratio (see Eq. \ref{eqtemp_ratio}).

We required the signal-to-noise ratio to be sufficient (even though the high-temperature lines are quite noisy) and the logarithm of the temperature to be between $log_{10}$(T) = 4.7 and 6. The lower limit is to avoid any optical thickness effects, and the upper limit is the highest temperature observable by SPICE during the observations.

A small selection of lines observed by SPICE can be used for temperature diagnostics via line ratios; Mg IX is an excellent candidate to measure the averaged corona around 1 MK \citep{del_zanna_solar_2018}, in combination with the  B-like Mg VIII, although the line from Mg VIII has a slight 
(20\%) density sensitivity. The ratio from C-like O III to B-like O IV could also be investigated, being dependent on temperature and very little on density. However, the temperature distribution seen in DEM is very broad: the temperatures extend from 30,000 K to 1 M K. Thus, the temperature value from the O IV/O III ratio would only represent one temperature value sampled from a very broad range, and not the temperature of the transition region plasma. 

\begin{figure}
\hspace*{-0.9cm}
     \begin{center}
            \includegraphics[width=9cm]{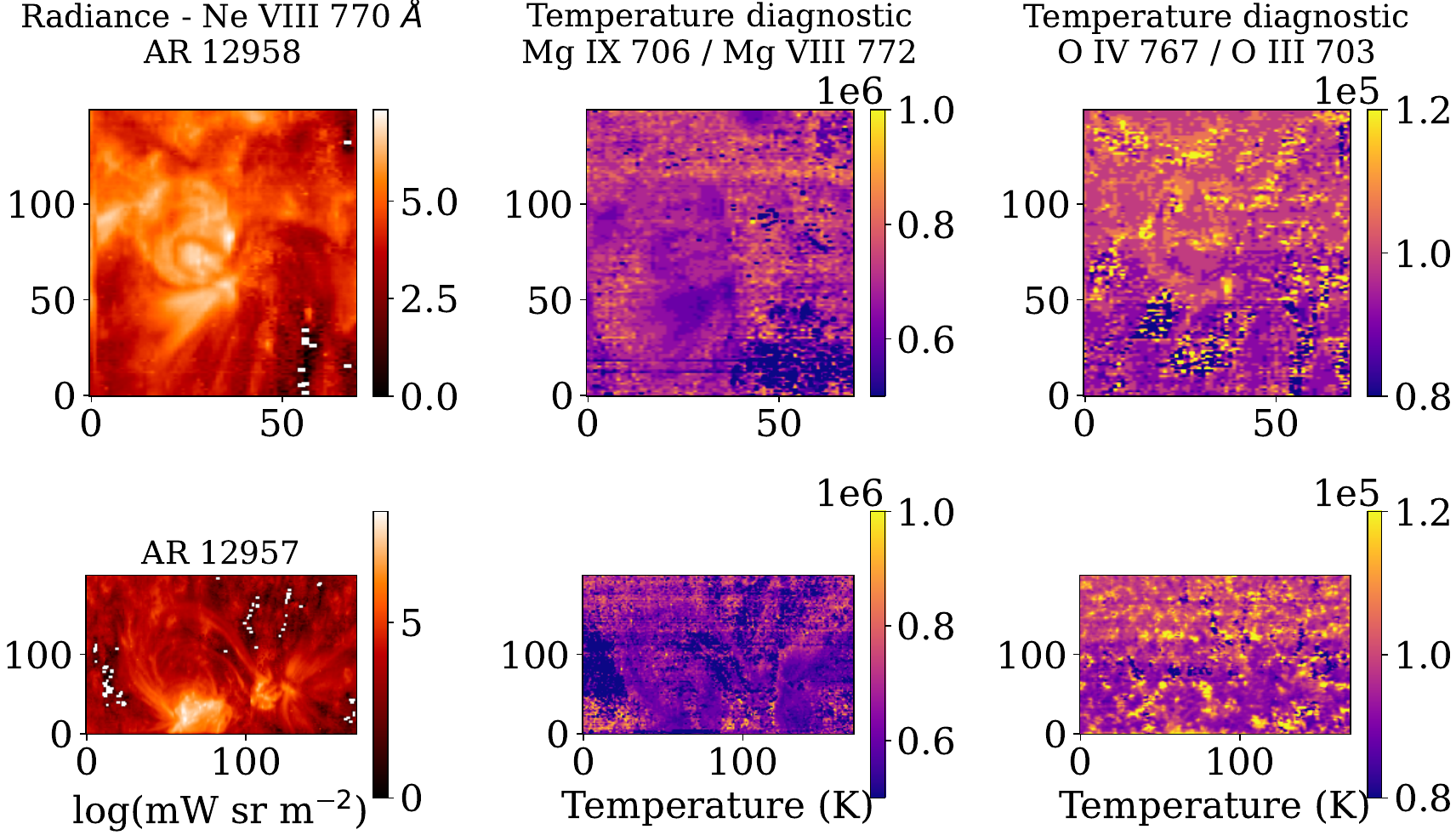}
    \end{center}
 \caption {Left panel: radiance map from the line Ne VIII (units in mW sr$^{-1}$ m$^{-2}$). \textcolor{black}{Middle and right panels: temperature diagnostic maps from element Mg and O.} Top row shows AR 12958, the bottom row shows AR 12957.}
 \label{fig:tempRatios}
 \end{figure}

\begin{figure}
    \centering
    \includegraphics[width=0.9\linewidth]{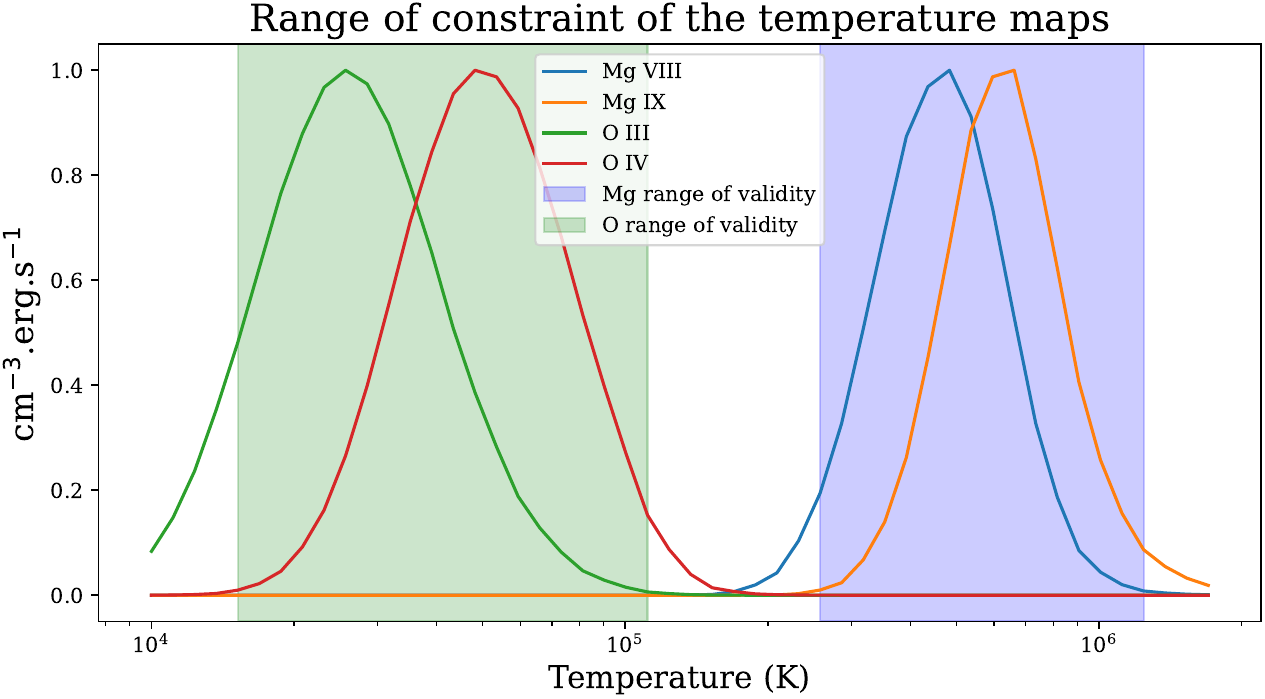}
    \caption{\textcolor{black}{Range of constraint of the temperature maps based on the contribution functions. The green shading represents the range of temperatures accessible with the oxygen line ratio diagnostic, and the purple shading the range of the magnesium ratio.}}
    \label{fig:constr_temp_maps}
\end{figure}

The line ratio method restricts the range of temperature solutions to $T_1 \leq T_0 \leq T_2$ ($T_1, T_2$ being the limits where the contribution functions of lines 1 and 2 are less than \textcolor{black}{1\% of their peak values \citep{2005ApJ...635L.101W}}). The range of constraint is presented in Figure \ref{fig:constr_temp_maps} for the magnesium and oxygen. 

The reliability of the magnesium ratio is limited to active regions, due to very low signal-to-noise ratio in quieter areas. 
On the middle panel of Figure \ref{fig:tempRatios}, the temperature diagnostic inferred from the two magnesium lines shows typical behavior, as a cooler temperature $T_0$ at the footpoints of the ARs and higher temperatures where upper transition region / coronal material is expected. Hotter regions are where we observe more coronal-like loops and brighter regions, as seen with EUI 174 \AA.
\textcolor{black}{The oxygen ratio has similar behavior, within its range of temperature. The chromospheric network is clearly distinguishable, however, these maps are still difficult to interpret; as well as for the Doppler maps, we \textcolor{black}{should} wait for the PSF correction to draw any conclusions.}

Three points are important to note: (1) those temperature maps are no more than a weighted average of the contribution functions and the emission measure; (2) the contribution functions inferred from the CHIANTI database overestimate the formation temperature of the ions involved here  \citep{dufresne2023benchmark}; 
(3) The accuracy of the temperature diagnostics relies on the reliability of the ionization/recombination rates and, most importantly, on the ionization equilibrium assumption. Those assumptions are acceptable in the corona but questionable in the transition region due to shorter dynamical timescales regarding recombination rates \citep{del_zanna_solar_2018}, especially around strong magnetic features. 
However, we find the temperature maps to be consistent with previous observations at these temperature ranges and they therefore provide useful information.\\\\

Throughout the last sections, we have given an overview of which analyses and diagnostics are likely to be most appropriate with SPICE data. These include radiance maps for distinct lines, intensity ratios for elemental composition, temperature diagnostics, and cross-analysis with other SolO instruments (EUI and HIS). While comparing elemental abundance data with other instruments is one of SPICE's main features, we can get a direct analysis of the abundance patterns by studying the FIP effect through SPICE-only data which we will do in the following section.

\section{Relative abundance diagnostics from line ratios and DEM methods}
\label{sectionFIP}

We define the FIP bias $f_X$ in the corona as
\begin{equation}
    f_X = \frac{Ab_X^C}{Ab_X^{Ph}},
\end{equation}
where $Ab_X^C$ and $Ab_X^{Ph}$ are the coronal and photospheric abundances of element $X$, respectively. SPICE observes lines from one low-FIP element (Mg), the intermediate-FIP element S, and the high-FIP elements N, O and Ne. Values of FIP bias around 1 are expected for coronal holes and sources of fast solar wind, while higher values (2-5) would indicate sources of slow solar wind and active regions. 

In selecting the lines which would be used for FIP bias computation, we have chosen those that best correspond to the four constraints defined by \cite{Feldman_2009} for Hinode/EIS lines, which are the following: 
\begin{itemize}
    \item consider only the lines with an upper atmosphere or coronal temperature ($\log(T) \gtrsim 5$),
    \item have a minimum brightness according to the CHIANTI database (Int $\geq 10^3$ erg cm$^{-2}$ sr$^{-1}$ s$^{-1}$), and no blended lines if possible,
    \item narrow down to lines with transitions between the first excited and ground configurations,
    \item and lastly, focus solely on lines with a low electron-density dependency. 
\end{itemize}
The chosen lines and their transitions are reported in bold in Table \ref{tab:compo_study}.

Typically, one tries to determine the relative FIP bias in the solar atmosphere of two elements from the ratio of the radiances of two spectral lines (one for each element).
The ratio of the FIP biases of these elements can be written in the coronal approximation \citep{Zambrana_Prado_2019} as:
\begin{equation}\label{eq:fipdem}
\frac{f_{X_{LF}}}{f_{X_{HF}}} = \frac{I_{LF}}{I_{HF}}
\left(
    \frac{Ab_{X_{LF}}^{Ph}}{Ab_{X_{HF}}^{Ph}} 
    \frac{\langle C_{LF}, \DEM\rangle}{\langle C_{HF}, \DEM\rangle}
\right)^{-1},
\end{equation}
where $I_\LF$ and $I_\HF$ are the radiances of the lines used for the diagnostics, low and high-FIP respectively; $C_\LF$ and $C_\HF$ are the corresponding contribution functions and $\langle a, b \rangle \equiv \int a(T) \, b(T) \; dT$.

All the atomic physics contributing to the emission of the ions is contained in the contribution functions. The DEM is tied to the distribution of plasma with temperature along the line-of-sight.

One then can do, for example, one of three following methods:
\begin{itemize}
    \item Find two lines with contribution functions similar enough that the ratio $\langle C_\LF, \DEM\rangle / \langle C_\HF, \DEM\rangle$ is constant, which we will call the 2LR (two-line ratio) method.
    \item Determine the DEM to compute the ratio of Equation \ref{eq:fipdem}, also called the DEM inversion method.
    \item Use a linear combination of spectral lines optimized for relative FIP bias determination. This is the linear combination ratio (LCR) method developed in \cite{Zambrana_Prado_2019}.
\end{itemize}
We will use these three techniques to determine the relative FIP bias in our observations.

\subsection{Two-line ratio method}

The 2LR method has been widely used for computing FIP bias maps \citep{2000A&A...364..835M}, but its problem resides in its reliance on very strong assumptions and hypotheses. In fact, any two lines involved must have very close formation temperatures and contribution functions -- to the extent that they can be approximated by the ratio of their maxima \citep{Zambrana_Prado_2019}. In real-world data, no two contribution functions are exactly the same; there is always at least a small dependence on the DEM. Besides that, unfortunately, the interesting lines recorded by SPICE for FIP diagnostic have very few similarities regarding their contribution functions \textcolor{black}{(see Appendix \ref{AppendixC})}.

\cite{1989ApJ...344.1046W} suggested the use of a simplification where the emission measure factor is canceled out by the ratio, and the relative FIP bias can simply be written as:
\begin{equation}\label{eq:2LReq}
    \frac{f_{\text{LF}}}{f_{\text{HF}}} = \frac{\max(C_{\text{HF}}) \, Ab_{\text{HF}}^{Ph}}{\max(C_{\text{LF}}) \, Ab_{\text{LF}}^{Ph}}\frac{I_{\text{LF}}}{I_{\text{HF}}},
\end{equation}
where $Ab_{X_{\text{LF/HF}}}^{Ph}$ is the (established) low-FIP / high-FIP photospheric abundance of element $X_i$, $\max(C_{X_i})$ is the maximum of the contribution function of the ion $X_i$ and $I_{X_i}$ is the observed intensity of element $X_i$.

The line pairs involving neon, magnesium, sulfur, oxygen, and nitrogen are thus the ones that received the most attention in this work. The Mg/Ne pairings are particularly interesting in the sense that the three ions (the two Mg ions and Ne) are formed at relatively high temperatures (respectively 0.8 MK, 1 MK and 0.63 MK), and are respectively low and high FIP elements (7.6 eV and 21.6 eV), which makes them very good candidates for meeting the assumptions of Section \ref{sectionFIP} \citep{2022ApJ...940...66B}. The FIP gap and close peak temperatures are the two main characteristics we look for when choosing a line pairing, but close contribution functions and good signal-to-noise ratio should also be taken into account. However, magnesium and neon do not have a good overlap temperature-wise (see Figures \ref{fig:loTcontrib} and \ref{fig:hiTcontrib}) and if not properly modeled, the measured FIP bias can be affected by the significant high-temperature tail of Ne VIII. The signal of the magnesium lines is also very weak outside the active regions\textcolor{black}{, especially for Mg VIII;} that is why the FIP bias measurements with these lines should only be trusted in the active regions.

While interesting, neon is a challenging element, since it has a noble gas configuration---hence its very high FIP---and is thus present in the photosphere mainly in its neutral form. As a result, no spectral lines for measuring the neon absolute abundance are available. 

\begin{figure*}
    \centering 
    {\includegraphics[width = \linewidth]{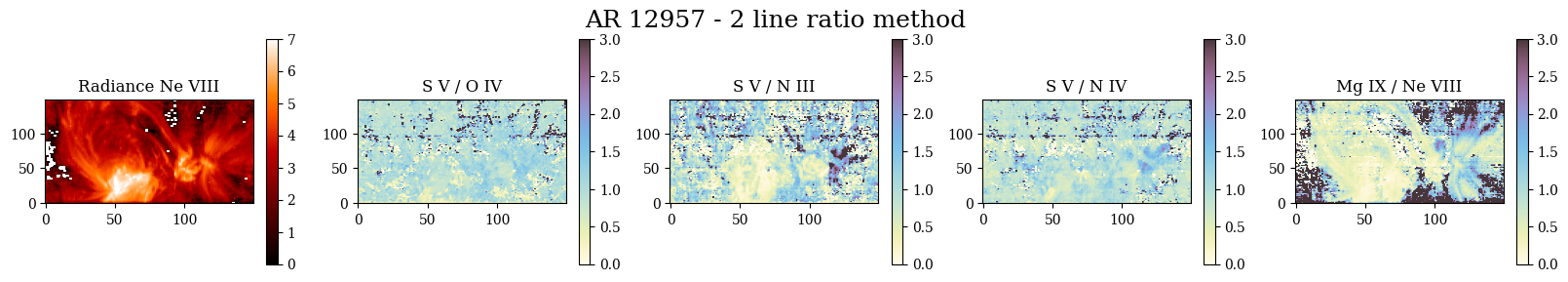}}
    \centering
    \includegraphics[width=\linewidth]{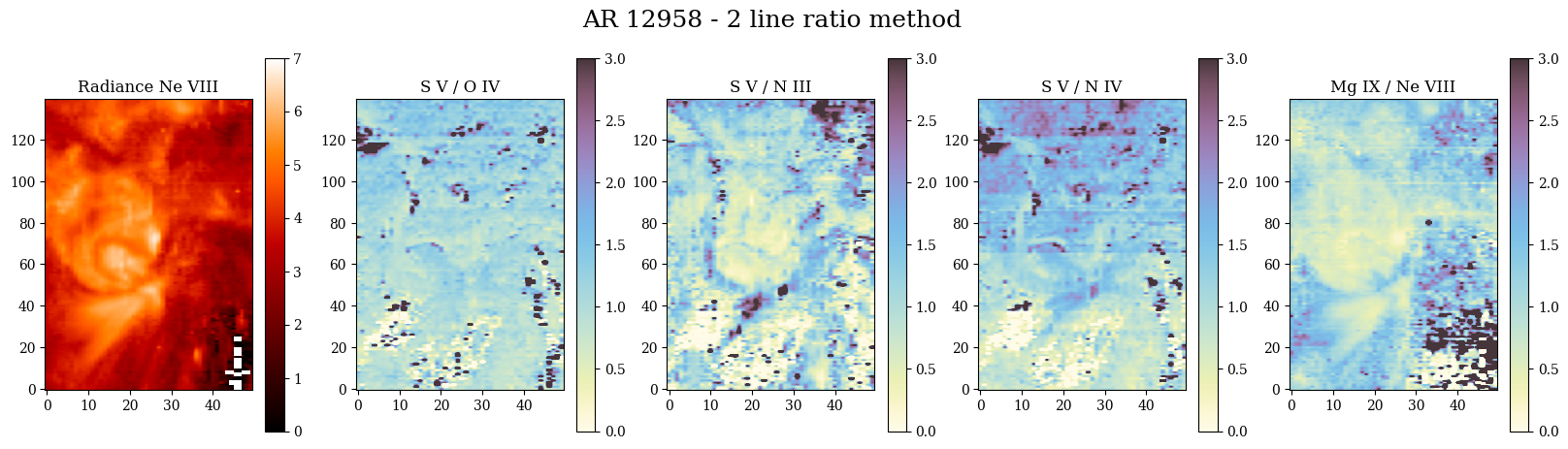}
    \caption{FIP bias diagnostic using the 2LR method. The top panel represents the active region observed in panel B of the mosaic, and the bottom panel is a zoom over the active region observed on panel A of the mosaic. White pixels are areas where the signal was too low to draw any conclusions on the diagnostic result. Darker areas suggest higher FIP-bias values, therefore coronal-like abundances and potential sources of slow solar wind.} 
    \label{fig:intensityratioNe}
\end{figure*}

Satisfying results have also been obtained with the pairing sulfur (low/intermediate-FIP) / nitrogen (high-FIP), and sulfur to oxygen even though sulfur is an intermediate FIP element. \textcolor{black}{While significant fractionation is observed with the S/N ratio, the S/O ratio does not present a strong discrimination. The behavior of the S/O ratio has been investigated in \cite{2020ApJ...895...36K}, and is a good candidate as the ratio is density insensitive between $1 \times 10^{7}$ and $1 \times 10^{10}$ cm$^{-3}$, and has a good match in temperature of formation. Consistent with their results, we found that the element sulfur is more abundant in open-field regions (see Fig. \ref{fig:intensityratioNe}, \ref{fig:FIPbiasDEM} and \ref{fig:FIP_20220302T18})}. SPICE also observes a wide variety of lines of those two elements, which leads to more consistent analysis.

Presented in Figure \ref{fig:intensityratioNe} are the results we obtained by applying the two-line ratio method (i.e., Equation \ref{eq:2LReq}) which gave the best results. The contribution functions were computed with an assumed electron density of $n_e = 2 \times 10^9$ cm$^{-3}$. The darker purple areas show low-FIP enhancement, so are a potential source of slow solar wind;  \textcolor{black}{black areas represent saturated FIP fractionation (which means, the exact value cannot be trusted but an enhancement of the low-FIP element is undeniable)}.

The magnesium and sulfur are highly enhanced in the regions where there is magnetic activity, especially at the footpoint of the AR 12957, which is consistent with the fact that previous observations from Hinode showed that edges of active regions are very good candidates for sources of slow solar wind, hence should show high relative FIP values.
Similar trends are observed with the ratio S/N, the latter having the advantage of being less sensitive to temperature -- in \textcolor{black}{contrast} to the ratio Mg/Ne, thermal structures are less distinct, due to closer formation temperatures.

The values of these line ratio maps are however to be taken with caution, as the 2LR method relies on many assumptions. The Widing and Feldman method can in fact 
produce incorrect results, if the plasma is not emitted
at the peak of the $C(T)$, which is often the case 
\citep[see discussion and references in][]{del_zanna_solar_2018}.
Moreover, as suggested by \cite{Warren_2016}, if physical mechanisms causing the FIP effect are averaged over time, which is the case with the relatively long SPICE exposures, the active regions might show a smaller value of FIP bias compared to older and long-lasting coronal structures.

\subsection{DEM inversion}

\begin{figure*} 
    \centering
    \includegraphics[width=\linewidth]{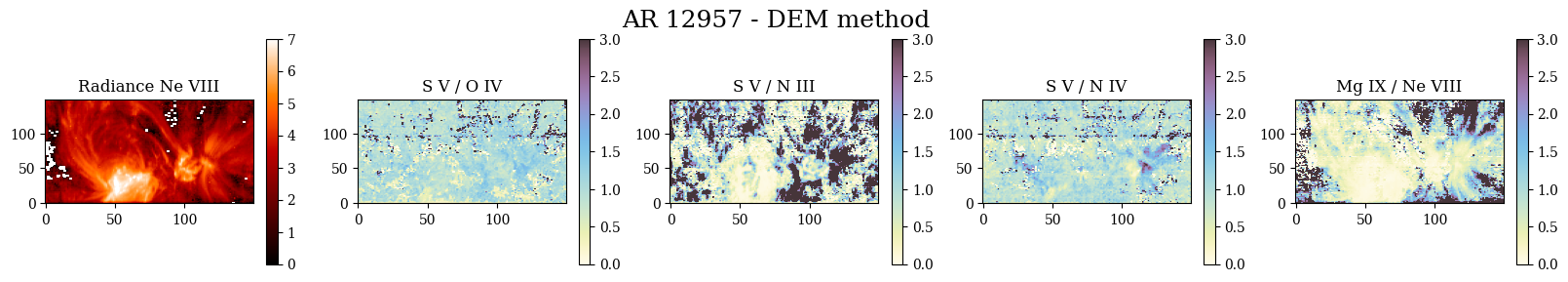}
    \includegraphics[width=\linewidth]{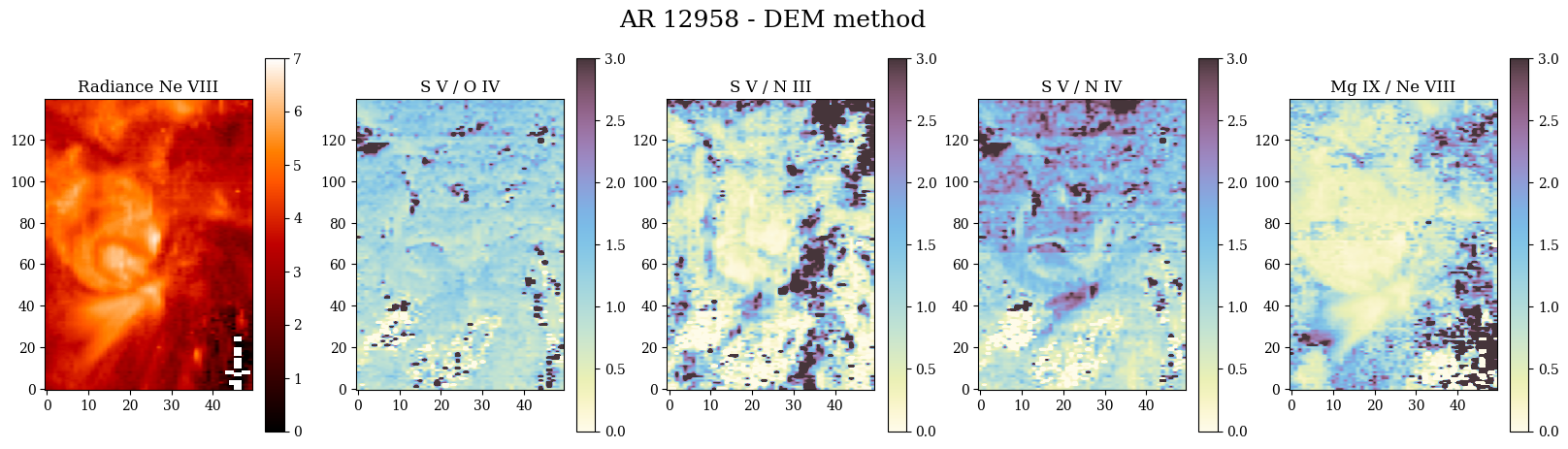}
    \caption{Relative FIP bias maps computed with the DEM method. Regions are the same as in Figure \ref{fig:intensityratioNe}. White pixels denote the same as in Figure \ref{fig:intensityratioNe}.}
    \label{fig:FIPbiasDEM}
\end{figure*}

We computed the relative FIP biases following Equation \ref{eq:DEMfip}, $<a,b>$ being the scalar product defined as $\sum_{i=1}^n a_ib_i$, $a, b \in \mathbb{R}^n$.
\begin{equation}\label{eq:DEMfip}
\centering
    \text{FIP}_{Bias} = \frac{I_{LF}}{I_{HF}}\left(\frac{A^{Ph}_{LF}}{A^{Ph}_{HF}}\frac{<C_{LF}, \DEM>}{<C_{HF}, \DEM>}\right)^{-1}
\end{equation}
The best result of the FIP bias values computed with the DEM is presented in Figure \ref{fig:FIPbiasDEM}, with the same layout as in Figure \ref{fig:intensityratioNe}. Thermal effects are still strong, despite the use of the DEM (the loop of AR 12957 is clearly distinguishable when using the neon and magnesium ratios, very likely due to the difference in their contribution functions (their peaks are more than 300 000 K apart for Mg IX / Ne VIII, 100 000 K apart for Mg VIII / Ne VIII). In spite of these thermal structures, we still observe a very good consistency with the 2LR method in the relative FIPs, namely higher values at the footpoints of the loops and certain uniformity in quiet areas.

A further improvement could be using temperature maps from other instruments such as Hinode/EIS to better constrain higher temperatures.

\subsection{Linear Combination Ratio}

\begin{figure}
    \centering
    \includegraphics[width=6cm]{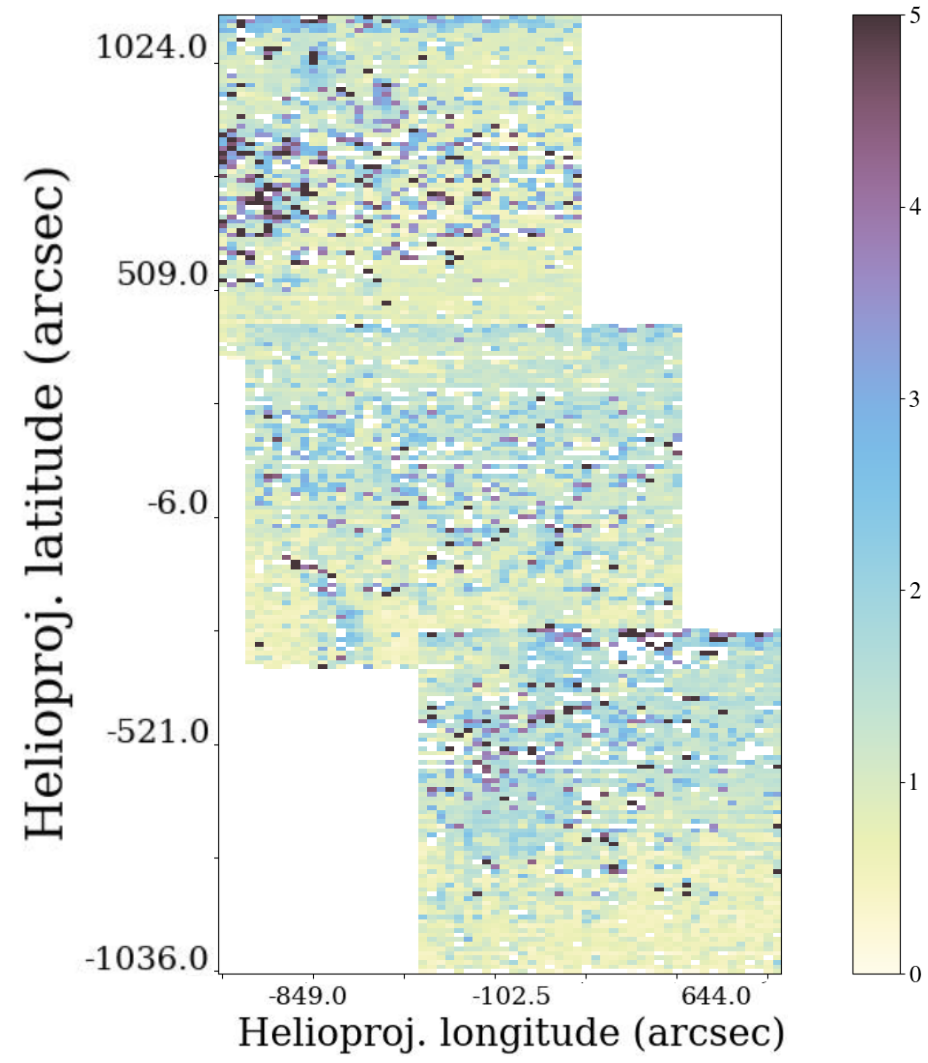}
    \caption{Relative FIP bias computed with the LCR method. The images have been binned by a factor of 3 along the X-axis and by a factor of 6 along the Y-axis. The values of the relative FIPs are as expected; around 1 in Quiet Sun areas, with higher values in active regions, especially around emerging areas.}
    \label{fig:FIP_20220302T18}
\end{figure}

We summarize here the principle of the Linear Combination Ratio method that is explained in detail in \cite{Zambrana_Prado_2019}.

Improving on the 2LR method, the LCR method uses two linear combinations of lines rather than only two lines. The resulting linear combinations of the contribution functions for low-FIP and high-FIP elements provide a better estimate of the relative FIP bias and render the method DEM-independent.
Using individual lines $i$ of low-FIP elements on one hand and $j$ of high-FIP elements on the other hand, we build linear combinations of their radiances $I_i$ from:
\begin{equation}
    \label{eq:pseudo_intensity}
    {I}_{LF} \equiv \sum_{i \in (LF)} \alpha_i \; \frac{I_i}{A^\phot_i}  \quad \text{and} \quad
    {I}_{HF} \equiv \sum_{j \in (HF)} \beta_j \; \frac{I_j}{A^\phot_j},
\end{equation}
where $\alpha_i$ or $\beta_j$ is a coefficient for line $i$ or $j$. These coefficients are optimized for the method to be as DEM-independent as possible (see details in \cite{Zambrana_Prado_2019} and \cite{Zambrana_Prado_2023}).
A Python module has been developed and is freely available at \url{https://git.ias.u-psud.fr/nzambran/fiplcr}. It can be used to compute the optimal coefficients $\alpha_i$ and $\beta_j$ to obtain accurate relative FIP bias maps from observations, without the need for computing the DEM.

Sulfur is considered to be a low FIP element for this diagnostic since no low FIP elements show spectral lines in the same temperature range as the selected \ion{N}{iii} and \ion{N}{iv} lines. Table \ref{tab:FIP_bias_ratios} lists the photospheric abundances we used to compute the relative FIP biases, taken from \cite{Asplund2021}. It lists as well the coronal abundances from \cite{Schmelz12} and the corresponding relative abundance bias which are only informative, to allow us to see in which areas we will obtain coronal abundances in our relative FIP bias maps. Table~\ref{tab:coefficients} shows the optimized coefficients obtained for each line and used in our calculations.

\begin{table}
\begin{center}
\begin{tabular}{lccccc}
\hline\hline
X & \mltc{FIP (eV)}   & $A^\coro_\mathrm{X}$      & $A^\phot_\mathrm{X}$  & $f_X$ & $f_\mathrm{S}/f_\mathrm{N}$  \\ \hline
S & 10.36             & $1.69\times 10^{-5}$      & $1.32\times 10^{-5}$  & 1.28  &  \multirow{2}{*}{1.42} \\ 
N & 14.03             & $5.17\times 10^{-5}$      & $6.76\times 10^{-5}$  & 0.78  &        \\ \hline \hline
\end{tabular}
\caption{First ionization potential of sulfur and nitrogen. }
\tablefoot{The coronal and photospheric abundances are taken from \cite{Schmelz12} and \cite{Asplund2021} respectively, their abundance biases, and the corresponding relative abundance bias.}
\label{tab:FIP_bias_ratios}
\end{center}
\end{table}

\begin{table}[]
    \centering
    \begin{tabular}{ccc}
    \hline \hline
    Spectral line   & Wavelength in $\AA$   & Coefficient   \\ \hline
    \ion{N}{iii}    & 991.577      & 1.0           \\
    \ion{N}{iv}     &765.152       & 2.0           \\
    \ion{S}{iv}     &750.221       & 3.2           \\
    \ion{S}{v}      &786.468        & 0.47          \\ \hline \hline
    \end{tabular}
    \caption{$\alpha$ and $\beta$ coefficients defined in Eq.~\ref{eq:pseudo_intensity} for each used spectral line, computed for $\log(n_e) = 9.5$.}
    \label{tab:coefficients}
\end{table}

For all three mosaic rasters, relative FIP bias maps were computed using the LCR method. They are depicted in Figure ~\ref{fig:FIP_20220302T18}. The white pixels correspond to pixels where the fitting of any of the four lines was not satisfying.
The obtained values of relative FIP bias peak near the photospheric value of reference, which is reassuring since the majority of the observed plasma is QS plasma and therefore is expected to have a mostly photospheric FIP bias. We see small areas of higher relative FIP bias, mainly around the footpoints of the loops of the ARs as well as in the center of the ARs, which is consistent with the fact that the abundance of low FIP elements is enhanced in areas where the magnetic field is closed. There is also an enhancement of the relative FIP bias in the fan loops in AR 12957.

Despite the binning of the spectral cube and the patchiness of the fitted radiance maps, which leads to
some white areas in the relative FIP bias maps, we find that the LCR FIP bias map obtained provides useful information on the FIP biases in the coronal structures in the field of view, which is remarkable given that it was produced without any DEM inversion and only four spectral lines.

\section{Discussion} \label{discussion}

SOOPs mosaics are designed to obtain a detailed map of the spatial distribution of solar atmospheric composition across an extended area of the Sun. One of their main goals is to link the remote sensing observations from Solar Orbiter with its in-situ data, especially with the SWA/HIS instrument, to link the in-situ events observed with their source regions on the Sun.
Using this SPICE mosaic, which is the first of its kind, we have been able to explore a range of analyses and diagnostic techniques that can be performed using such spectroscopic raster data, as well as performing some preliminary processing to set standards for future observations.

Despite working with the best available calibrated data, we encountered several problems while working with SPICE data, particularly concerning dark subtraction. The pipeline incorporates an automated algorithm designed to select the most appropriate dataset for dark current subtraction. By identifying the dataset closest in time to the observations, this algorithm generally yields highly accurate outcomes. However, it has come to our attention that the algorithm encountered difficulties in accurately matching the data during the period of 2-6 March 2022. This discrepancy can be attributed to temperature fluctuations in the instrument's detector that occurred on 1-2 March. Notably, these composition maps generated on March 2, 2022, employed lengthy exposures of 120 s, exacerbating the issue compared to shorter exposures.

The impact of a mismatched dark subtraction is intricate in nature. While it is expected to have a negligible effect on strong or bright lines, its influence on weak or faint lines will be proportionally greater. As a result, the observed line ratios can be significantly altered as a consequence of this disparity. The degradation of the \textcolor{black}{detectors (burn-in)} makes this survey an important milestone as the data recorded cannot be reproduced, or with an increasing degradation factor.

Despite those issues, radiance maps show satisfying agreement with their higher resolution counterparts taken from EUI, not only in active and hot regions but also in quieter areas and coronal holes, where the visual texture of these regions matches between the two instruments. This result is very encouraging for future observations, and reinforces SPICE's ability to track detailed solar structures; despite the spatial binning we currently employ. We believe we have clearly demonstrated the capability of SPICE observations to complement in-situ and other observations. 
Regarding the lines available, some blends are still unresolved (e.g., S IV 750 \AA\;and Mg IX 749.5 \AA), and we are limited by the small number of high-temperature lines (peaking around 1 MK). Having more of those higher-temperature lines could enable us to perform more accurate temperature and density diagnostics by adding more constraints to the problem, typically in DEM inversion. Another issue concerning the DEM calculation is that fixed photospheric abundances are assumed in the computation process. But if coronal plasma is observed along the line of sight, the lines are brighter than they would be with photospheric observations, therefore the model will tend to underestimate the line emission in those lines. It will try to compensate for this incoherence by increasing the EM. This effect is, however, limited by constraints applied by other lines that are not affected by this issue, and it happens that it is the case for most lines used for DEM computation. The overall effect we observe is that the DEM is higher in certain temperature ranges, to compensate for the weakest lines in the model. 

A way to address this issue would be to iterate the DEM computation, starting with photospheric abundances and adjusting them at each iteration until convergence. We could use EUI 174 \AA\;data to add a high-temperature constraint (the 174 \AA\;channel is dominated by Fe X, log(T)=6.1), and possibly STIX data. 


Even though retrieving HIS data is still a work in progress because of calibration issues of the instrument, in-flight anomalies and complicated quality evaluation due to SSD noise signatures, we plan in future works to compare in-situ and remote-sensing data and to recover the relative elemental abundances of common observed elements as well as their speed. The purpose of merging these two types of data is to perform back-mapping of heliospheric mechanisms. We seek changes in the chemical compositions, but also in velocity and ionization level measurements. Sulfur might be the key to searching for open fields \citep{2018SoPh..293...47R}.
As pointed out in \cite{2019ApJ...879..124L} and in \cite{2020ApJ...895...36K}, intermediate-FIP elements as sulfur behave as high-FIP elements in closed magnetic loops, but behave as low-FIP elements in the solar wind. Moreover, sulfur fractionates only in the low chromosphere where nonresonant waves on open field lines occur. 
Resonant Alfvèn waves can produce sufficient fractionation only in the top part of the chromosphere, since they have most of their energy trapped at higher altitudes, on top of the loops. In these conditions, the fractionation will not be significant, which is what we observe in our S/O ratio.
The interpretation of the behavior of sulfur in remote-sensing observations might be more complex and has to be carefully considered, but will be a crucial variable to understand the solar wind mechanisms.

The combination of HMI and EUV spectroscopy data can provide a more complete understanding of the dynamics and energetics of the solar atmosphere. The comparison of HMI magnetograms with EUV emission images can aid in associating magnetic structures to chromospheric and coronal features, such as flares, prominences, and coronal loops. By tracking the temporal and spatial evolution of these features, we can investigate the underlying physical mechanisms that govern the heating and acceleration of the plasma in the solar atmosphere. 

Regarding temperature diagnostics, the results are to be taken with caution. First of all, the coronal assumption that is being made is a strong hypothesis and results in significant uncertainties in the results. Moreover, models for the CHIANTI database are still being improved, and it has been found that the formation temperature of transition region ions is probably overestimated in many cases. Strong lines in the EUV also present some anomalous ions (especially from Li-like and Na-like ions, in our case O VI and Ne VIII), which present additional uncertainties and a strong density dependence. Single-line ratios accuracy and issues are discussed in detail in \cite{2001A&A...379..708D} and \cite{2003A&A...406.1089D}.

Another aspect that is still to be corrected is the PSF, and doing so will enable us to compute more precise and accurate Doppler maps \citep{plowman2022spice}. This will provide additional information about sources of solar wind, as a significant portion of the slow solar wind would form in upflowing, blue-shifted plasma. 
The most promising aspect of these composition map analyses is the relative FIP bias maps, especially performed with the Linear Combination Ratio. Our maps show good consistency with previous publications and will be the key to understanding the formation and source regions of the solar wind.

\section{Conclusion}

We present an overview of possible diagnostics techniques for analyzing SPICE composition observations, and more precisely for the connection mosaic performed in March 2022. Mapping the source regions on the surface of the Sun to the heliosphere using these connection mosaics is an important objective of Solar Orbiter and is planned to be performed periodically during future Solar Orbiter remote sensing windows.

From this preliminary mosaic, we have obtained encouraging results by reducing the data and extracting a range of quantities, and we explored the range of what SPICE data has to offer. 

In addition to radiance maps, we have computed Differential Emission Measures, elemental abundances, and relative FIP bias maps from the SPICE data. This illustrates that SPICE has a wide array of outputs to provide and could be even more formidable if paired with \textit{in-situ} analyses. Proposals for joint analysis have already been explored \citep{2018IAUS..335...87F}, and Solar Orbiter measurements are very promising for addressing the key questions of solar-terrestrial physics.
Radiances inferred from this dataset show variances acceptable for active regions but quiet Sun discrepancies still need to be explained. More observations will complete the preliminary diagnostic performed in this work. 
The study of the radiance maps has shown that SPICE can capture complex solar structures with satisfying resolution (even with a factor of two binning in the slit direction for all maps). Isothermal temperatures (around logT = 5.8 / 5.9) have been found for AR loop structures, with an enhancement of low-FIP elements of a factor 3-4. Structures in SPICE radiance maps can be easily associated with those seen in EUI observations, enabling a clear correspondence that allows other SPICE quantities, such as FIP bias, to be oriented relative to those higher-resolution observations. 

After investigating three different methods of deriving FIP-bias maps, it would seem that the LCR with NIII, NIV, SIV, SV works best (Fig \ref{fig:FIP_20220302T18}) – showing high FIP values at the base of some AR structures, (footpoints of loops and fan loops), \textcolor{black}{considered to be potential source regions of the solar wind. This method seems to be less sensitive to noise and low signal, as it shows very steady values for the more quiet regions. We also notice a significant change in the behavior of sulfur, following the magnetic field geometry.}

Expanding this analysis to include observations from other EUV instruments, such as Hinode/EIS will enable a broader temperature coverage, extending investigations to the higher corona.

In the future, we will utilize combined measurements from SWA/HIS and Hinode/EIS to conduct comprehensive analyses of the data and validate the initial findings presented in this study.  
These observations will not only provide comprehensive diagnostics from the chromosphere to the upper corona but also ensure temporal coverage of the specific regions under investigation.

In summary, SPICE offers the ability to measure composition at a wide range of heights and temperatures in the solar atmosphere, from the chromosphere through the transition region to the low corona, complementing the observations and data from Hinode/EIS and SOHO/SUMER. Comparing these observations with observations from SWA/HIS allows us to map back the source regions at the surface to the solar wind. Specifically, these SPICE measurements show higher FIP-bias values to be present at the footpoints of coronal loops associated with two active regions. These measurements also demonstrate that FIP fractionation can occur at lower heights in the solar atmosphere (upper chromosphere and transition region), as shown by the behavior of the intermediate-FIP element Sulfur.

\begin{acknowledgements}
These efforts at SwRI for Solar Orbiter SPICE are supported by NASA under GSFC subcontract \#80GSFC20C0053 to Southwest Research Institute. The development of the SPICE instrument has been funded by ESA member states and ESA (contract no. SOL.S.ASTR.CON. 00070). The work at GSFC is supported by NASA funding for Solar Orbiter SPICE, and N. Zambrana-Prado is supported by cooperative agreement 80NSSC21M0180. The German contribution to SPICE is funded by the Bundesministerium für Wirtschaft und Technologie through the Deutsches Zentrum für Luft und Raumfahrt e.V. (DLR), grants no. 50 OT 1001/1201/1901. The Swiss hardware contribution was funded through PRODEX by the Swiss Space Office (SSO). The UK hardware contribution was funded by the UK Space Agency. S.L. Yardley would like to thank STFC via the consolidated grant (STFC ST/V000497/1). G. Del Zanna acknowledges support from STFC (UK) via the consolidated grant to the atomic astrophysics group at DAMTP, University of Cambridge (ST/T000481/1).

Python modules used for this work include \texttt{numpy}, \texttt{matplotlib}, \texttt{astropy}, \texttt{sunpy}, \texttt{scipy}, \texttt{spice\_uncertainties}\footnote{\url{https://git.ias.u-psud.fr/spice/spice\_uncertainties/-/tree/main}}, \texttt{EMToolKiT}\footnote{\url{https://github.com/jeplowman/EMToolKit}} and \texttt{cblind}\footnote{\url{https://github.com/volodia99/cblind}}.

\end{acknowledgements}

\bibliographystyle{aa}
\bibliography{bibliography.bib}

\begin{appendix} 
\section{Line fitting and comments on the windows}\label{AppendixA}

\begin{figure}[h!]
    \begin{minipage}[t]{.45\linewidth}
        \centering \includegraphics[width=\linewidth]{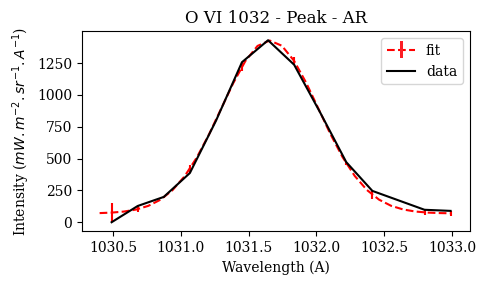}
   \end{minipage} \hfill
   \begin{minipage}[t]{.45\linewidth}
        \centering \includegraphics[width=\linewidth]{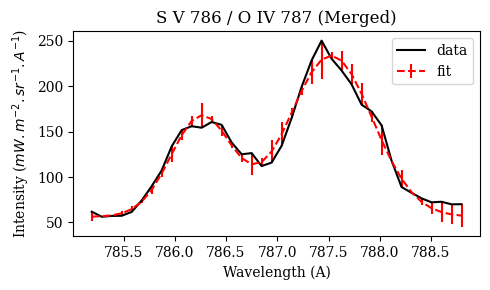}
    \end{minipage} \hfill
   \begin{minipage}[b]{.45\linewidth}
        \centering \includegraphics[width=\linewidth]{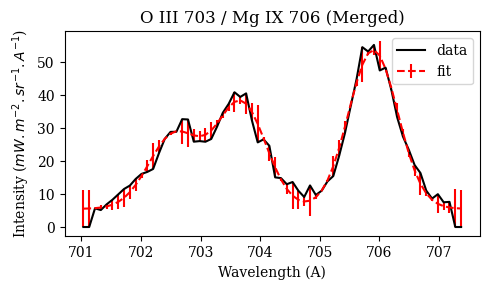}
    \end{minipage} \hfill
   \begin{minipage}[b]{.45\linewidth}
        \centering \includegraphics[width=\linewidth]{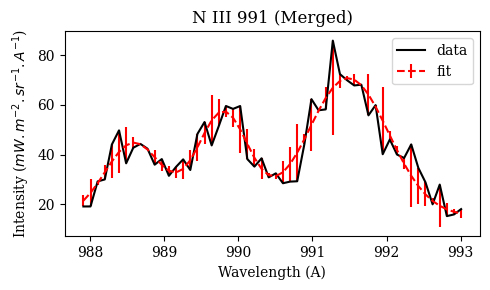}
    \end{minipage}
\caption{Examples of fits for single, double and triple Gaussian models over the AR 12958.}
\label{fig:fitgauss}
\end{figure}

In order to properly analyze the spectrum (intensity, shift, turbulence), a fit is needed. We rely on the \texttt{specutils} \citep{nicholas_earl_2022_6207491} module from the \cite{astropy:2013}. 
We have chosen to model the spectrum with a sum of Gaussian curves and a constant to fit both the peaks and the continuum.
We require an initial guess for each of the parameters of the Gaussian (here, the mean $\mu$, the standard deviation $\sigma$ and the amplitude $A$). The guesses are made using the astropy \texttt{estimate\_line\_parameters}, which relies on a least-squares fit to the input data. If more than one peak is fitted, we divide the spectrum in sub-spectra to compute those initial guesses, as indicated in Table \ref{tab:compo_study}. Once the parameters for each Gaussian are obtained, we compute the final fit using the whole spectrum.
The results of the single-peak fit are presented in Fig. \ref{fig:fitgauss}.

The main source of noise in SPICE files is shot noise, also called Poisson noise, which tends to dominate at lower photon counts.
 The model of errors that will be used in the following analysis is the sum (the noise is additive) of the errors induced by the dark (Poissonian), background (Poissonian), read (Gaussian) and line (Poissonian) noises. For each pixel, these noise sources add in quadrature, as 
\begin{equation}
    N_{total} = \sqrt{{N_{dark}^2} + N_{background}^2 + N_{read}^2 + N_{line}^2}
\end{equation}
A noise floor is also initialized based on SPICE's uncertainty package, developed by E. Buchlin. Its value is equal to the sum of dark current and background signal levels.

The goodness of fit is computed as the residuals between the model predictions and the observed data points. For this non-linear least squares fitting, the goal is to minimize the sum of squares of these residuals. The goodness of fit is evaluated as the mean of the difference between the observed data points and the corresponding values predicted by the model, sliced with the same windows we use for estimating the Gaussian parameters. This goodness of fit can be negligible regarding the instrumental error (especially for lines N IV and O VI, which are single-peaked).

\FloatBarrier

\subsection{O III 703 / Mg IX 706}
This window raises an additional challenge because of the O III multiplet around 703 \AA. Indeed, within the range [702.3 - 703.8], there are 6 O III lines and two non-negligible S III lines. This multiplicity of lines makes the fitting much more difficult because one fitted peak will not correspond to a single ion. That is why the O III results should be taken with great caution, as we do not know exactly the contribution of each line. To have greater accuracy and more reliable results, we chose (despite a higher computing time) to fit a model involving three Gaussians, even though each Gaussian will fit a blend of several ions. We focused on two main oxygen lines (O III 702.8 -- highly blended with S III 702.7 / 702.8 and O III 703.8) and the Mg IX 706 line. 

As with all blended windows, the contribution of the strongest line is often non-negligible when doing a line fitting, even though two peaks are clearly distinguishable. If one wanted to be more exact with the continuum fitting, instead of modeling it by a constant, one could model a linear function for the magnesium emission peak, which is tilted because of the contribution of the strong O III line.

\subsection{Ne VIII 770 / Mg VIII 772}
The same problem as mentioned above is encountered with this window, in addition to the fact that the magnesium VIII is barely detectable outside of very bright and hot active regions. As seen on Figure \ref{fig:avg_spectra}, the quiet Sun measurements of Mg VIII can not be reliable; however, within an active region, we have sufficient signal, even though it stays fairly weak.  

\subsection{S IV 748 / S IV 750}
\begin{figure*}[h!]
    \centering
    \includegraphics[width=\linewidth]{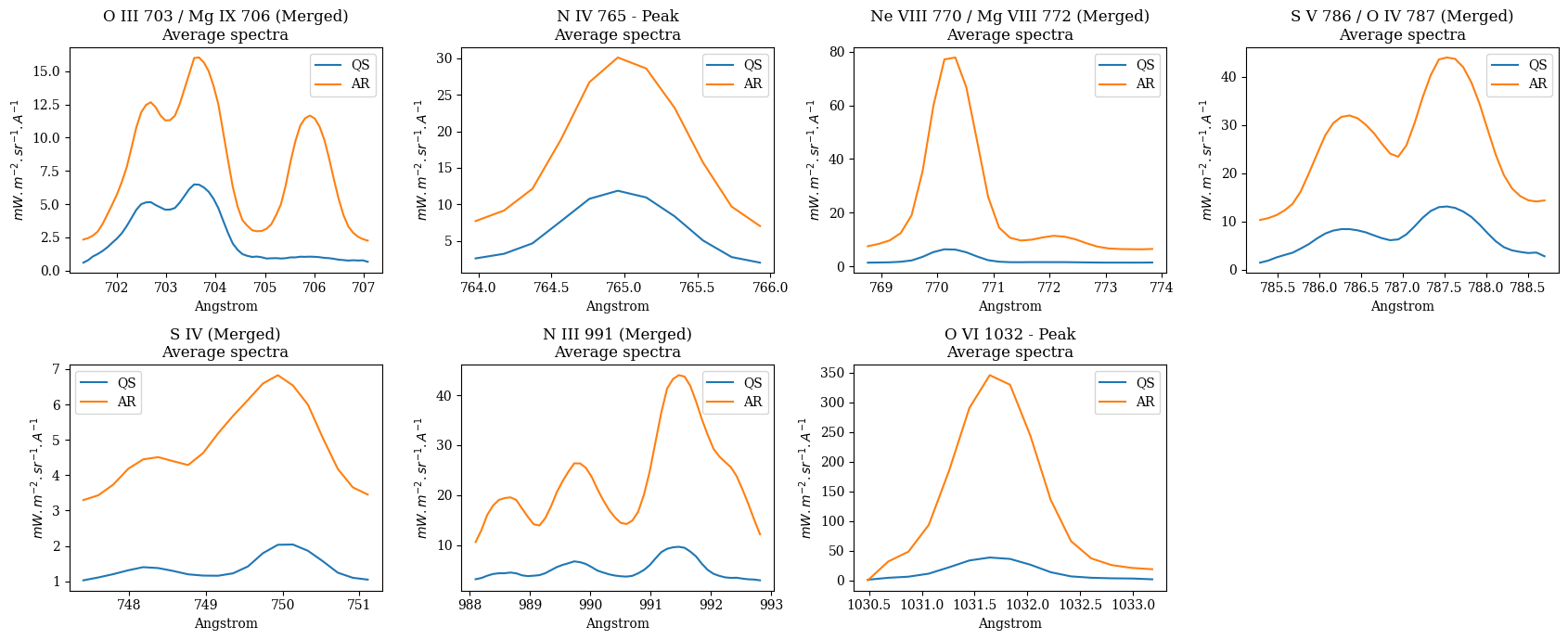}
    \caption{Averaged spectra over the cubes of the \textcolor{black}{raster "A"} of the mosaic. The blue curve represents the spectra averaged over a QS area, the orange one the spectra averaged over an AR area. The AR and QS areas correspond to S1 and S2, respectively, on Figure \ref{fig:S1S2}.}
    \label{fig:avg_spectra}
\end{figure*}

As can be seen on Figure \ref{fig:avg_spectra}, the spectra of this spectral window look less reliable than others, with what looks like a bad dark subtraction coupled with very wide peaks. The asymmetry on the second peak above 750 \AA\;is very likely due to a blend with the Mg IX 749.5 \AA. This might result in an overestimation of the sulfur line in the diagnostics -- it can be seen on Figure \ref{fig:totrad2} that AR structures are brighter in the S IV 750 \AA\;line.

\subsection{N III 991}
The N III window raises the same concern as the O III / Mg IX window. We also chose a model with 3 Gaussians to maximize our accuracy. We focused on the peak at 988.6 \AA, which may be a blend of O I and Na VI, the peak at 989.8 \AA\;accounting for N III and the N III 991.5 \AA\;peak. See Figure \ref{fig:fitgauss} for an example of the fits.
An issue that we encountered was the presence of cosmic rays, which the fitting algorithm frequently misidentified as emission peaks. To address this issue, we imposed constraints on the standard deviation during the generation of the Gaussian model. While this measure did not entirely resolve the problem, it successfully eliminated the majority of cosmic rays.

The maps obtained are shown in Figures \ref{fig:totrad} and \ref{fig:totrad2}. The windows Ne VIII / Mg VIII, N IV - Peak and O VI - Peak present a binning on the spectral axis, which requires a multiplication of the radiances by a factor of two.

\FloatBarrier

\section{Contribution functions} \label{AppendixC}

\FloatBarrier
\begin{figure}[h!]
    \centering \includegraphics[width=8cm]{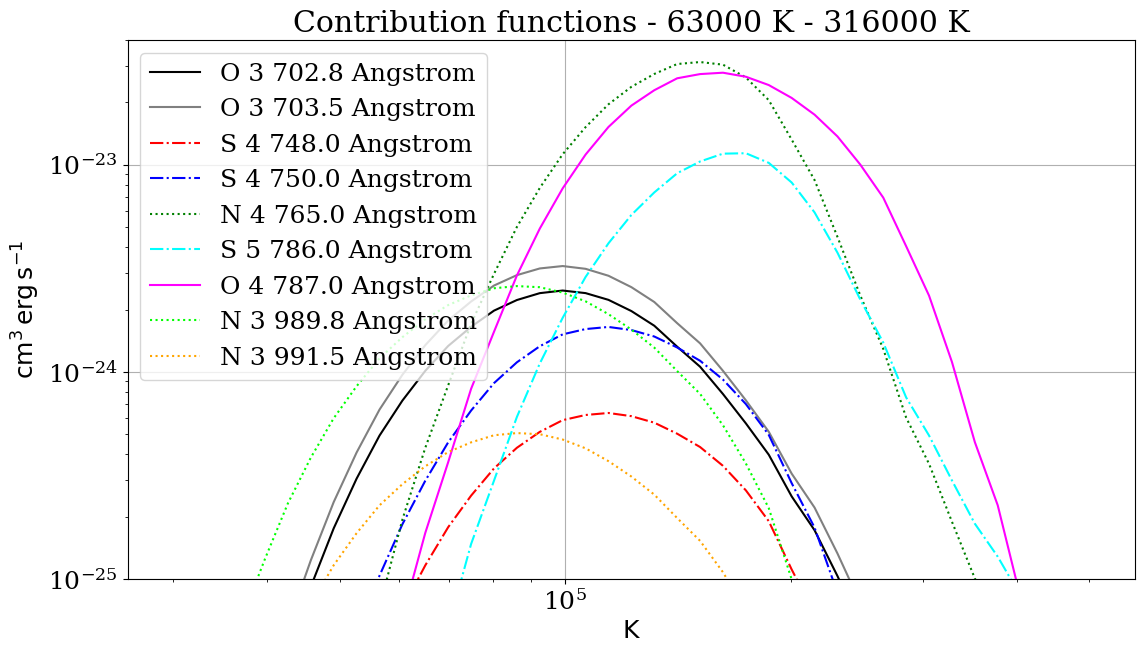}
    \centering \includegraphics[width=8cm]{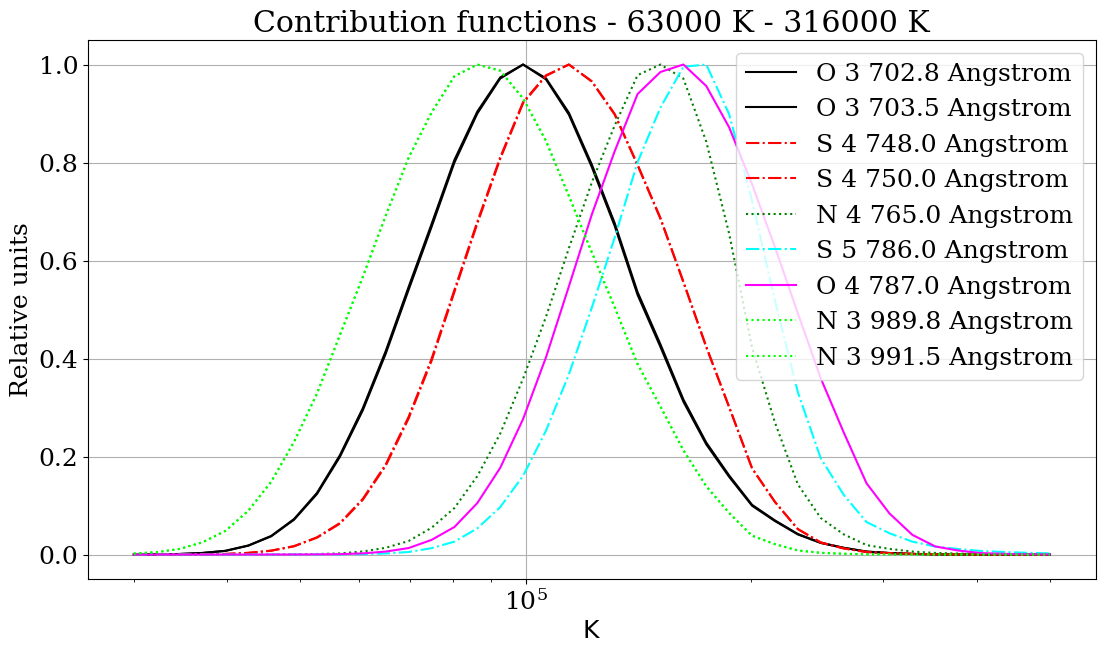}
    \caption{Contribution functions for lower temperature lines - $6.3\times 10^4$ - $3.16\times 10^5$ K. Dotted lines represent the nitrogen contribution functions and dashed-dotted ones the sulfur lines. Top: contribution functions displayed on a log-log scale. \\Bottom: contribution functions normalized and displayed on a linear y-scale. Lines overlapping in temperature have been plotted in the same color for less confusion. }\label{fig:loTcontrib}
\end{figure}
\FloatBarrier
\begin{figure}[h!]
    \centering \includegraphics[width=8.2cm]{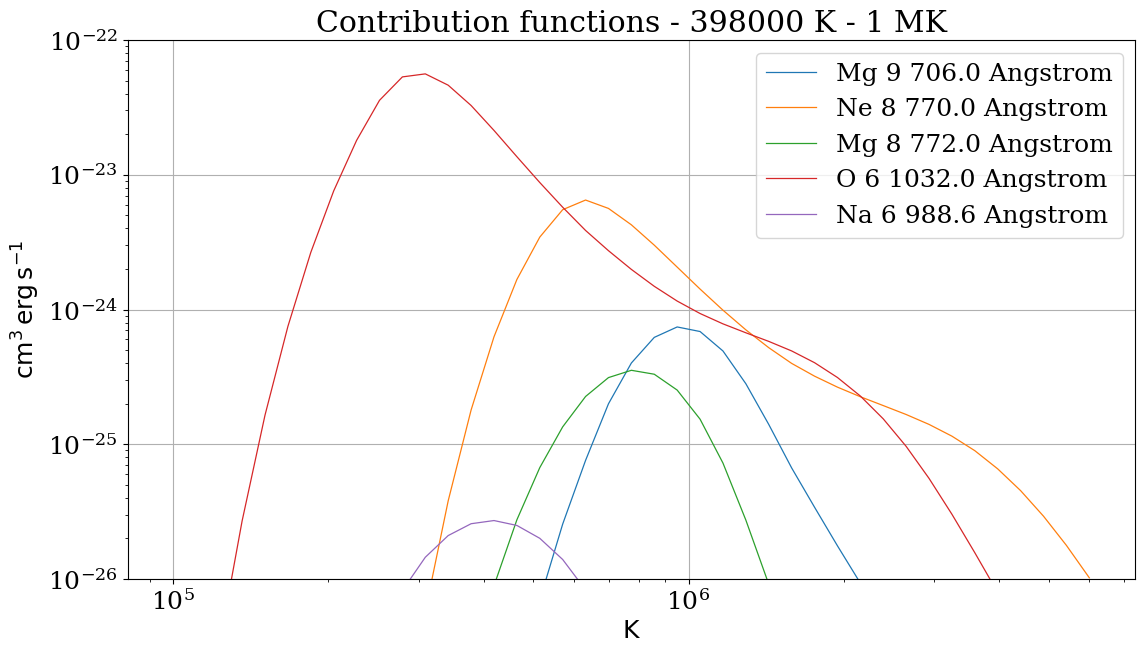}
    \centering \includegraphics[width=8.2cm]{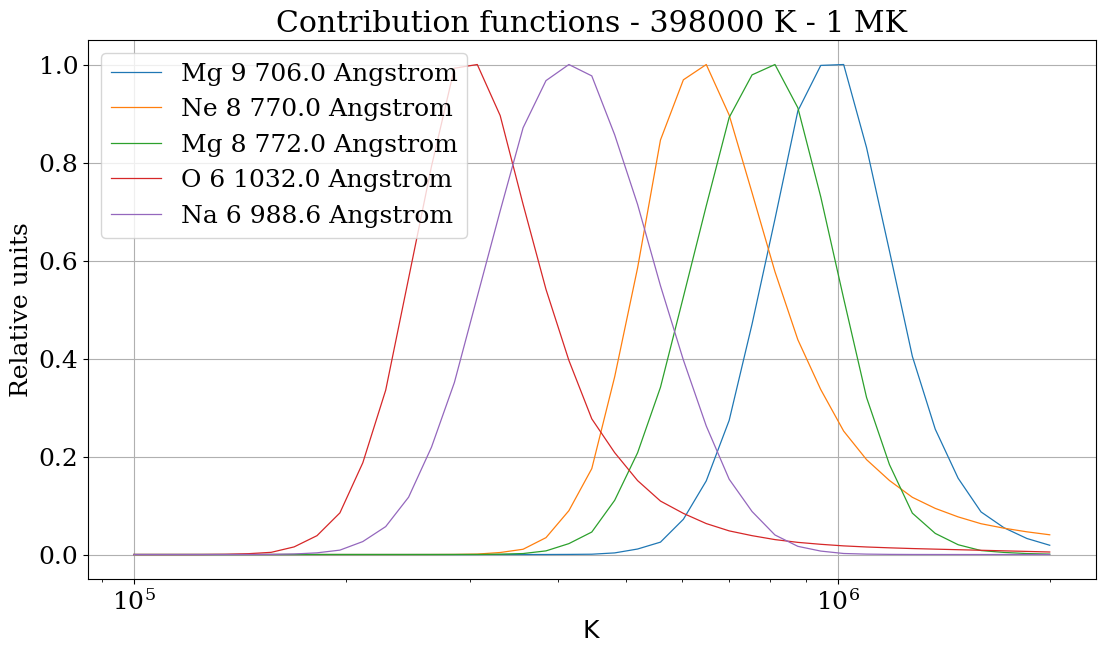}
    \caption{Contribution functions for higher temperature lines - $3.9\times 10^5$ - $1.0\times 10^6$ K. Top and bottom panels are the same as Figure \ref{fig:loTcontrib}.}
    \label{fig:hiTcontrib}
\end{figure}
\FloatBarrier

As can be seen on Figures \ref{fig:loTcontrib} and \ref{fig:hiTcontrib}, where contribution functions $C_{ij}$ are plotted with an electron density of $n_e=2\times 10^{-9}$ cm$^{-3}$, only a few couples of lines satisfy the condition of having "close enough" contribution functions. Among those, the best pairs would be S V / O IV, S V / N IV, S IV / O III and undoubtedly S IV 748 / S IV 750. For higher temperatures, the contribution functions are more sparse. However, the couples Mg IX / Ne VIII, Mg VIII / Ne VIII and Mg IX / Mg VIII would still be usable.

\section{Radiance maps}\label{fullrad}

 \begin{figure*}[]
    \centering
    \includegraphics[width=15cm]{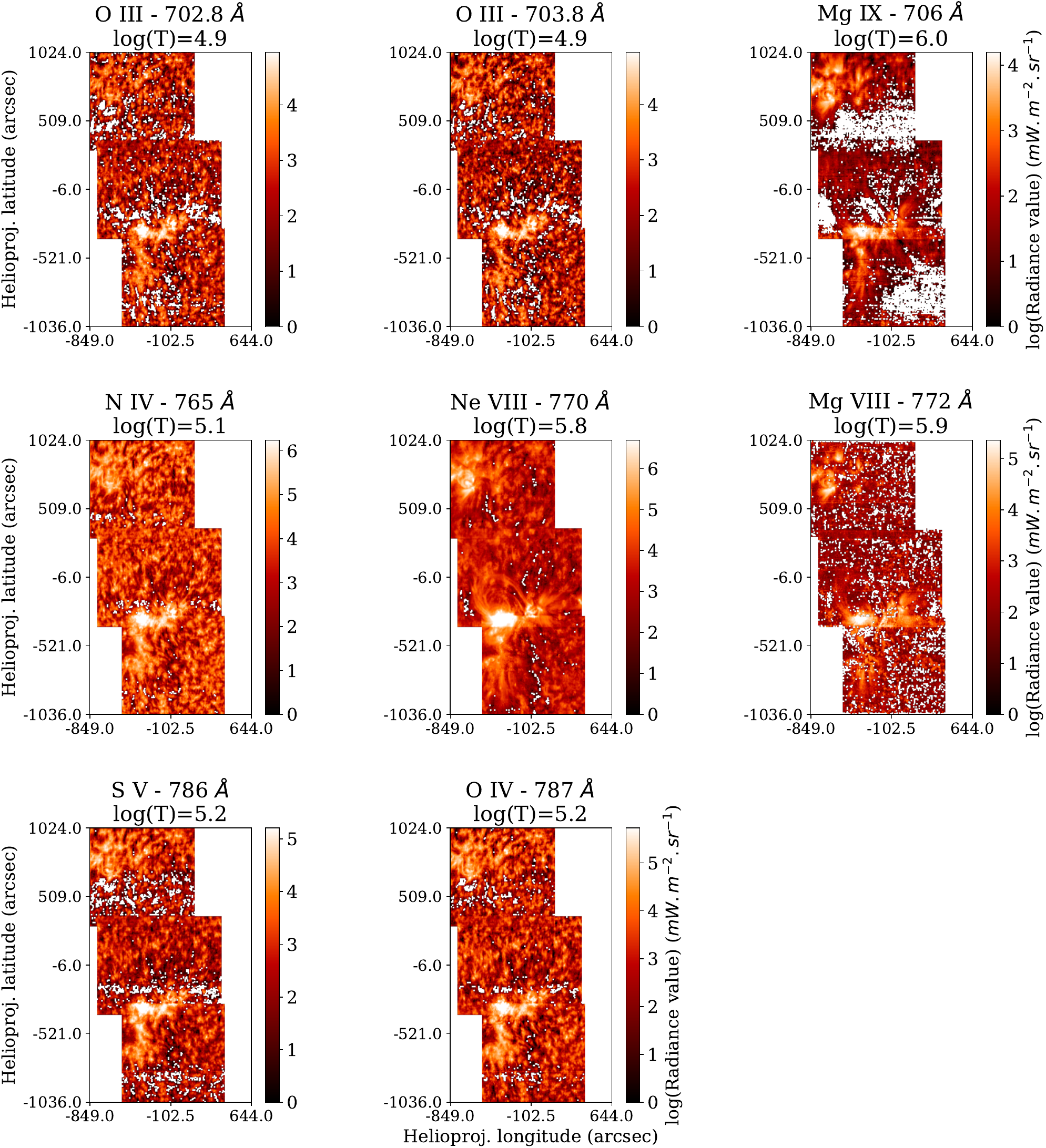}
    \caption{Radiances maps from the lines recorded by SPICE during the SOOP Mosaic on March 2$^{nd}$, 2022. Units are in  mW sr$^{-1}$ m$^{-2}$.}
    \label{fig:totrad}
\end{figure*}  

\begin{figure*}[]
    \centering
    \includegraphics[width=15cm]{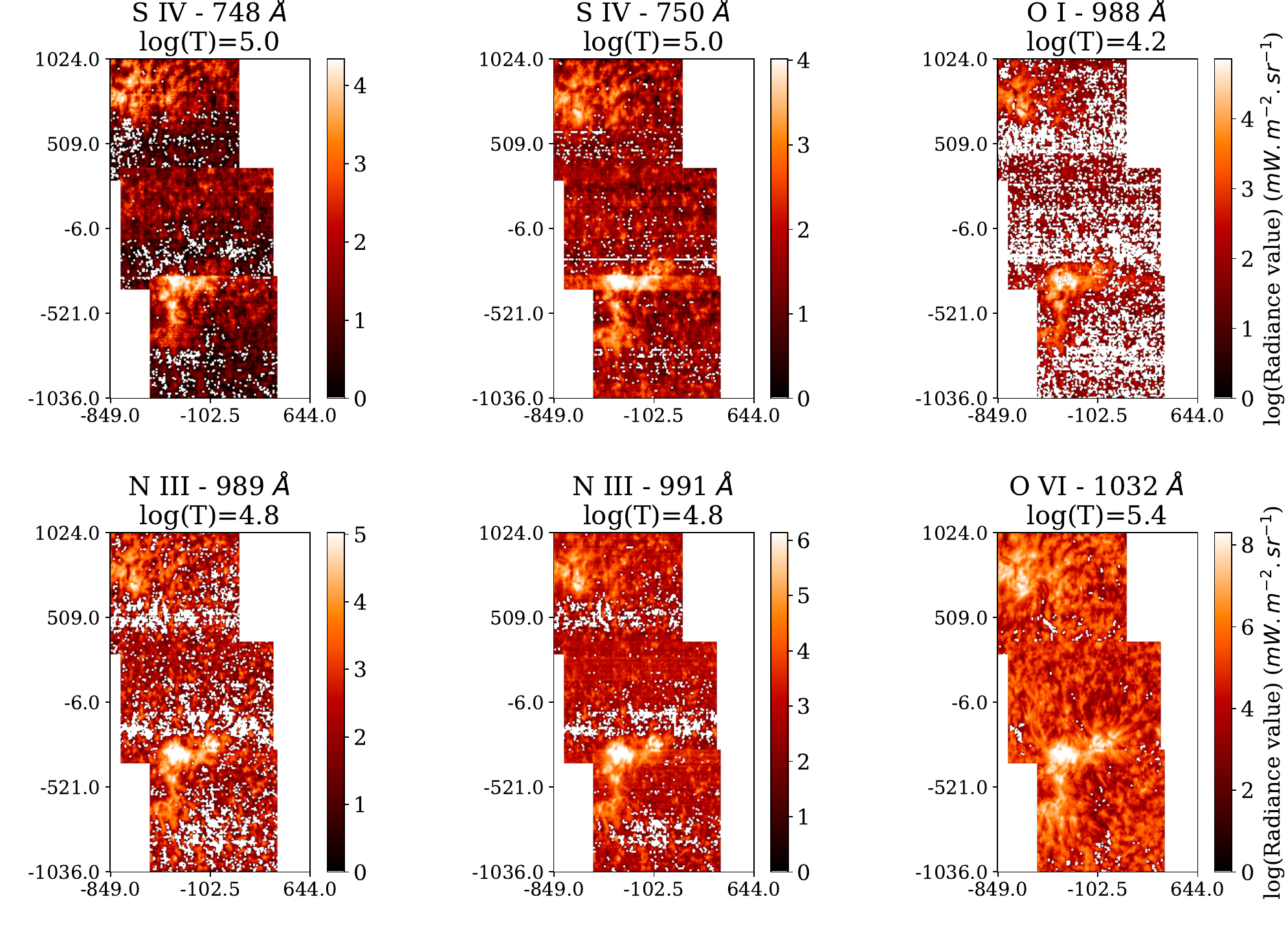}
    \caption{Radiances maps from the lines recorded by SPICE during the SOOP Mosaic on March 2$^{nd}$, 2022. Same units as in Figure \ref{fig:totrad}.}
    \label{fig:totrad2}
\end{figure*}     

\end{appendix}

\end{document}